\documentclass[11pt]{article}
\usepackage{jheppub}

\usepackage{amsmath}
\usepackage{amssymb}
\usepackage{verbatim}

\usepackage[T1,OT1]{fontenc}

\usepackage{graphicx}

\allowdisplaybreaks[3]

\usepackage[hang]{footmisc}
\newcommand{\e}{\mathrm{e}}
\newcommand{\w}{\wedge}

\newcommand{\nl}{\notag \\ &\quad\,}
\newcommand{\nll}{\notag \\ &}

\title{LVS de Sitter Vacua are probably in the Swampland}

\author{Daniel Junghans}

\affiliation{Harvard University, Center of Mathematical Sciences and Applications, 20 Garden Street,\\Cambridge, MA 02138, USA}

\emailAdd{djunghans @ cmsa.fas.harvard.edu}

\notoc

\abstract{
We argue that dS vacua in the LARGE-volume scenario of type IIB string theory are vulnerable to various unsuppressed curvature, warping and $g_s$ corrections.
We work out in general how these corrections affect the moduli vevs, the vacuum energy and the moduli masses
in the 4D EFT for the two K\"ahler moduli, the conifold modulus and a nilpotent superfield describing the anti-brane uplift.
Our analysis reveals that the corrections are parametrically larger in the relevant expressions than one might have guessed from their suppression in the off-shell potential.
Some corrections appear without any parametric suppression at all, which makes them particularly dangerous for candidate dS vacua. Other types of corrections can in principle be made small for appropriate parameter choices. However, we show in an explicit model that this is never possible for all corrections at the same time when the vacuum energy is positive.
Some of the corrections we consider are also relevant for the stability of non-supersymmetric AdS vacua.
}

\begin{document}

\numberwithin{equation}{section}

\maketitle

\newpage

\section{Introduction}

The LARGE-volume scenario (LVS) \cite{Balasubramanian:2005zx, Conlon:2005ki, Cicoli:2007xp, Cicoli:2008va} is
one of the leading proposals for moduli stabilization in Calabi-Yau (CY) orientifold compactifications of type IIB string theory.
Its central claim is the existence of non-supersymmetric AdS vacua at exponentially large volumes, which are constructed using a combination of perturbative and non-perturbative corrections to the classical scalar potential.
It was furthermore argued that the
AdS vacua can be uplifted to meta-stable dS vacua by adding a suitable source of additional energy such as an anti-D3 brane (as originally proposed in the KKLT scenario \cite{Kachru:2003aw}). Although this procedure is technically rather involved, models realizing the LVS have become increasingly explicit in the last years. In particular, the interesting work \cite{Crino:2020qwk} managed to construct LVS dS vacua in an explicit CY model including
a conifold region with an anti-brane.

On the other hand, it was conjectured in the context of the swampland program that dS vacua \cite{Brennan:2017rbf, Danielsson:2018ztv, Obied:2018sgi}
and non-supersymmetric AdS vacua \cite{Ooguri:2016pdq, Freivogel:2016qwc, Danielsson:2016mtx} are in general inconsistent in quantum gravity. One may therefore wonder whether string-theory effects not taken into account in the LVS could invalidate some of its claims.

The goal of this paper is to perform a detailed analysis of this question. In particular, we study various types of corrections to the LVS potential in a 4D EFT including the two K\"ahler moduli, the conifold modulus and a nilpotent superfield describing the anti-brane uplift.
While several earlier works already studied corrections to the LVS potential \cite{Conlon:2005ki, Cicoli:2007xp, Cicoli:2008va, Berg:2007wt, Conlon:2010ji}\footnote{See also \cite{Cicoli:2021rub} for a recent general discussion of $\alpha^\prime$ corrections in IIB/F-theory compactifications.},
we go beyond these results
in several ways. First, we consider a more general setup including the conifold modulus and nilpotent superfield, which allows us to study more explicitly the dS uplift in the LVS. Previous analyses of corrections did not include these two fields and focussed on the robustness of the AdS vacua.
Second, we work out a number of corrections to the potential that have not been considered in the context of the LVS before.
Finally, we derive model-independent, analytic expressions for the corrected moduli vevs, vacuum energy and moduli masses. These expressions reveal previously unnoticed control problems.

A key issue we identify is that the corrections are parametrically less suppressed in the relevant expressions than one might naively conclude from the off-shell potential. This phenomenon can be traced back to a $1/g_s$ scaling in the vev of the small K\"ahler modulus and a related cancellation effect we call the non-perturbative no-scale structure (NPNS).
As a consequence, some types of corrections are not suppressed by any small parameters
but rather blow up at small coupling, leading to large uncertainties in the moduli vevs, the vacuum energy and the moduli masses.
While one might hope to avoid this problem by searching for models where these corrections are absent, we show that the remaining types of corrections are dangerous as well. The parametric dependence of the latter is such that they can in principle be made small by suitable parameter choices. However, interestingly, this is not possible for all corrections at the same time. Indeed, we show in the explicit model of \cite{Crino:2020qwk} that, assuming $\mathcal{O}(1)$ numerical coefficients, at least one of these corrections becomes large at every point in the parameter space where the vacuum energy is positive. Our results thus suggest that it is in principle impossible to construct reliable dS vacua in the LVS such that all corrections can be self-consistently neglected.\footnote{It was recently argued that loop corrections below the KK scale can make the cosmological constant of LVS AdS vacua positive in the IR \cite{deAlwis:2021zab}. We will not study this scenario in this work.}

This paper is organized as follows. In Section \ref{setup}, we review the LVS potential including the dS uplift. In Section \ref{vol}, we discuss a number of possibly dangerous corrections to the LVS potential. In Section \ref{npns}, we explain a cancellation effect in the vacuum energy and the moduli masses we call the NPNS. We then analyze in Section \ref{problems} how the various corrections affect the properties of LVS vacua. In particular, we state analytic expressions for the moduli vevs, the vacuum energy and the moduli masses.
In Section \ref{dsmin}, we study the corrections in the explicit model of \cite{Crino:2020qwk}.
We conclude in Section \ref{concl} with a discussion of our results.

\section{LVS Potential}
\label{setup}

We now briefly review the LVS potential including the dS uplift, mostly following \cite{Crino:2020qwk}. For a more detailed discussion, we refer to \cite{Crino:2020qwk} and the earlier works \cite{Balasubramanian:2005zx, Conlon:2005ki, Cicoli:2007xp, Cicoli:2008va}. For simplicity, we restrict to the case of a Swiss-cheese CY $X$ with $h^{1,1}(X)=2$ and an orientifold projection such that $h_-^{1,1}(X)=0$.

Our starting point is the low-energy EFT of the two K\"ahler moduli $T_b$ and $T_s$ and the conifold modulus $Z$.
The anti-brane uplift is incorporated through the presence of an additional nilpotent superfield $Y$ (see, e.g., \cite{Kallosh:2015nia} and references therein). We denote by $\tau_b=\text{Re}(T_b)$ and $\tau_s=\text{Re}(T_s)$ the (Einstein-frame) volumes of the two 4-cycles as usual. Furthermore, $\zeta=|Z|$ parametrizes the deformation of the conifold region \cite{Klebanov:2000hb, Giddings:2001yu}. The potential for $Z$ in the strongly warped regime was derived in \cite{Douglas:2007tu, Douglas:2008jx, Bena:2018fqc, Blumenhagen:2019qcg, Dudas:2019pls}.

After the axio-dilaton and all complex-structure moduli except for the conifold modulus are integrated out, the effective K\"ahler potential and superpotential for the remaining fields are\footnote{The standard procedure is to simply freeze the axio-dilaton and the complex-structure moduli at constant vevs rather than integrating them out properly (i.e., treating them as functions of $T_i$, $Z$). One may wonder whether this approximation misses $T_i$ and $Z$ dependent corrections to the effective $\mathcal{K}$ and $W$.\vspace{\maxdimen} Neglecting $Z$, one can show that the axio-dilaton and the complex-structure moduli approximately decouple from the $T_i$ in the LVS such that corrections to the frozen vevs are $1/\mathcal{V}$ suppressed \cite{Achucarro:2008sy, Gallego:2008qi}. Including the $Z$ terms in $\mathcal{K}$ and $W$, we checked in explicit examples that the suppression factor can be $\gg 1/\mathcal{V}$ but is typically still small so that the freezing approximation remains valid when $\mathcal{K}$ and $W$ are given by \eqref{k0}, \eqref{w0}.
}
\begin{align}
\mathcal{K} &= -2 \ln \left(\mathcal{V} + \frac{\xi}{2g_s^{3/2}}\right) + \gamma_0\left(\frac{Y\bar Y}{\mathcal{V}^{2/3}} + \frac{c^\prime \xi^\prime |Z|^{2/3}}{\mathcal{V}^{2/3}}\right)-\ln\left(\frac{2}{g_s}\right)+\text{const.}, \label{k0} \\
W &= W_0 + A_s \e^{-a_s T_s} - \frac{M}{2\pi i}Z\left(\ln \frac{Z}{\Lambda_0^3} -1\right)+\frac{iK}{g_s} Z + \frac{i\sqrt{c^{\prime\prime}}}{\sqrt{\pi}g_sM} Z^{2/3} Y. \label{w0}
\end{align}
Here, $W_0$ is the flux superpotential
(excluding the $Z$ terms), $g_s$ is the string coupling and $A_s$, $a_s$ are numbers depending on the non-perturbative effect
\cite{Witten:1996bn, Kachru:2003aw}
on the small divisor. $K$ and $M$ are (in our convention positive) flux numbers characterizing the conifold region \cite{Klebanov:2000hb, Giddings:2001yu}. $\gamma_0$ and $\Lambda_0$ are model-dependent constants discussed below. The (Einstein-frame) CY volume is defined as
\begin{equation}
\mathcal{V} = \tau_b^{3/2}-\kappa_s \tau_s^{3/2}, \label{ve}
\end{equation}
where $\kappa_s$ is a constant related to the triple-intersection numbers and we absorbed an analogous constant in front of $\tau_b$ into the definition of the latter. We furthermore have
\begin{equation}
c^\prime\approx 1.18, \qquad c^{\prime\prime}\approx 1.75, \qquad \xi= -\frac{\chi(X)\zeta(3)}{2(2\pi)^3}, \qquad \xi^\prime = 9g_sM^2, \label{constants}
\end{equation}
where $\zeta(3)\approx 1.20$ and $\chi(X)$ is the Euler number of the CY.

Before we move on, let us make a few comments about the $Z$, $Y$ dependence of \eqref{k0} and \eqref{w0}.
First, we only kept in $\mathcal{K}$ the $Z$ term which is leading near the solutions we will consider, i.e., in the regime
$\zeta\ll 1$.\footnote{It was recently shown in \cite{Bento:2021nbb} that there are also dS solutions in a less strongly warped regime where a second term $\sim|Z|^2\ln\frac{|Z|}{\Lambda_0^3}$ becomes relevant in $\mathcal{K}$. We will not consider such solutions in this paper, but it would be interesting to see whether they are also affected by our arguments.}
Second, $\gamma_0$ and $\Lambda_0$ are model-dependent constants which would in principle have to be determined by gluing the conifold region to the CY bulk.\footnote{In particular, $\gamma_0$ has to be determined by correctly normalizing $\mathcal{V}$ as the volume of the glued CY.}
In the following, we will set $\gamma_0=\Lambda_0=1$ for concreteness, as implicitly assumed in \cite{Crino:2020qwk} and other works.
Choosing different values for $\gamma_0$ and $\Lambda_0$ does not qualitatively influence the conclusions of this paper but may affect the numbers computed in Section \ref{dsmin}.
Finally, the derivation of the $|Z|^{2/3}$ term in \eqref{k0} and the $Z^{2/3} Y$ term in \eqref{w0}
assumes that the off-shell $Z$ dependence of the warp factor in the conifold region is $\e^{-4A}\sim |Z|^{-4/3}$ \cite{Douglas:2007tu, Douglas:2008jx, Bena:2018fqc, Blumenhagen:2019qcg, Dudas:2019pls}. However, it is not clear whether this assumption is justified \cite{Gao:2020xqh} since from the Klebanov-Strassler solution \cite{Klebanov:2000hb} one can only read off the \emph{on-shell} warp factor $\e^{-4A}\sim |Z_0|^{-4/3}$, where $Z_0$ denotes the value at which $Z$ is stabilized.
Since we do not have anything new to say about this issue, we will follow \cite{Douglas:2007tu, Douglas:2008jx, Bena:2018fqc, Blumenhagen:2019qcg, Dudas:2019pls, Crino:2020qwk} and proceed under the assumption that $\mathcal{K}$ and $W$ take the above form.

Using \eqref{k0} and \eqref{w0}, we can now compute the $F$-term scalar potential. After integrating out $Y$ and the axionic parts of $T_b$, $T_s$ and $Z$, we obtain a potential for the three real scalars $\tau_b$, $\tau_s$ and $\zeta$:
\begin{align}
V &= \frac{4a_s^2|A_s|^2g_s\sqrt{\tau_s}\e^{-2a_s\tau_s}}{3\kappa_s \mathcal{V}} - \frac{2a_s|A_s|g_s\tau_s|W_0|\e^{-a_s\tau_s}}{\mathcal{V}^2} + \frac{3|W_0|^2\xi}{8\sqrt{g_s}\mathcal{V}^3} \nl + \frac{\zeta^{4/3}}{2\pi^2 c^\prime \mathcal{V}^{4/3}} \left( \frac{\pi c^\prime c^{\prime\prime}}{g_sM^2} + \frac{\pi^2 K^2}{g_s^2M^2} + \frac{\pi K}{g_sM}\ln\zeta + \frac{1}{4}\ln^2\zeta \right), \label{lvs-potential}
\end{align}
where we only displayed the leading terms.
Note that the terms in the first line correspond to the usual two-moduli LVS potential without uplift (yielding non-supersymmetric AdS vacua), while the second line contains the leading $\zeta$ terms including the uplift term $\sim c^{\prime\prime}$ generated in the presence of an anti-D3 brane.

Minimizing the potential, one finds the solution
\begin{align}
\mathcal{V} &= \frac{a_s\tau_s-1}{4a_s\tau_s-1}\frac{3\kappa_s|W_0| \sqrt{\tau_s}}{a_s|A_s|}\, \e^{a_s\tau_s}, \label{tb0} \\
\tau_s &= \frac{\xi^{2/3}}{(2\kappa_s)^{2/3}g_s} + \frac{1}{3a_s} + \frac{4\alpha}{15a_s} + \mathcal{O}(g_s), \label{ts0} \\
\zeta &= \e^{-\frac{2\pi K}{g_sM}-\frac{3}{4}+\sqrt{\frac{9}{16}-\frac{4\pi c^{\prime}c^{\prime\prime}}{g_sM^2}}}. \label{zeta0}
\end{align}
Because of the square root in the last line, the solution requires
\begin{equation}
g_sM^2\ge \frac{64\pi c^{\prime}c^{\prime\prime}}{9}\approx 46.1. \label{gsm}
\end{equation}
This is the conifold-instability bound derived in \cite{Bena:2018fqc, Blumenhagen:2019qcg}.

The $\alpha$ term in \eqref{ts0} captures the backreaction of the anti-D3 uplift on $\tau_s$ (with $\alpha=0$ in the absence of anti-D3 branes and $\alpha>0$ otherwise).
Explicitly, $\alpha$ is defined as
\begin{equation}
\alpha = \frac{20a_s q_0 \zeta^{4/3} \mathcal{V}^{5/3}}{27g_s|W_0|^2\kappa_s \sqrt{\tau_s}} \label{alpha}
\end{equation}
with
\begin{equation}
q_0 = \frac{3}{32\pi^2 c^\prime} \left(3-\sqrt{9-\frac{64\pi c^\prime c^{\prime\prime}}{g_sM^2}}\right). \label{q0}
\end{equation}

One checks that evaluating \eqref{lvs-potential} at the minimum yields
\begin{equation}
V_0 = \frac{3g_s\kappa_s|W_0|^2\sqrt{\tau_s}}{8\mathcal{V}^3a_s}\left( \alpha-1+ \mathcal{O}(g_s)\right), \label{v0}
\end{equation}
where we denote by $V_0$ the on-shell value of $V$.\footnote{Here and in the following, we avoid putting a 0 index on moduli vevs in order to not clutter the equations.} At small $g_s$, we thus require $\alpha>1$ for a dS solution.

A point that will become important below is that $V_0$ is parametrically smaller than the terms in the off-shell potential \eqref{lvs-potential} by a factor $g_s$.\footnote{This was also pointed out in \cite{Hebecker:2012aw}.} Indeed, using $\tau_s\sim\mathcal{O}(1)/g_s$ in $V_0$, one finds $V_0 \sim \sqrt{g_s}|W_0|^2/\mathcal{V}^3$. On the other hand, the individual terms in the first line of  \eqref{lvs-potential} are of the order $V \sim |W_0|^2/\sqrt{g_s}\mathcal{V}^3$ near the minimum. As will be explained in Section \ref{npns}, this is due to a cancellation effect in the LVS potential which we call the non-perturbative no-scale structure (NPNS).

In order to check the stability of the dS solutions, we also need to know the eigenvalues of the mass matrix.
The kinetic terms for the complex fields $\Phi^i = (T_s, T_b, Z)$ are given by $\mathcal{L}_\text{kin}= - K_{i\bar\jmath} (\partial_\mu \Phi^i) (\partial^\mu \bar\Phi^{\bar\jmath})$ as usual.
Using this, we find $\mathcal{L}_\text{kin} \supset -Q_{ij} (\partial_\mu\phi^i) (\partial^\mu\phi^j)$ with $\phi^i =(\tau_s,\tau_b,\zeta)$ and
\begin{equation}
Q=\begin{pmatrix}
\frac{3\kappa_s}{8\sqrt{\tau_s}\mathcal{V}} & -\frac{9\kappa_s\sqrt{\tau_s}}{8\mathcal{V}^{5/3}} & \frac{3 c^\prime g_sM^2\kappa_s\sqrt{\tau_s}}{2\zeta^{1/3}\mathcal{V}^{5/3}} \\
-\frac{9\kappa_s\sqrt{\tau_s}}{8\mathcal{V}^{5/3}} & \frac{3}{4\mathcal{V}^{4/3}} & -\frac{3c^\prime g_sM^2}{2\zeta^{1/3}\mathcal{V}^{4/3}}  \\
\frac{3c^\prime g_sM^2 \kappa_s\sqrt{\tau_s}}{2\zeta^{1/3}\mathcal{V}^{5/3}} & -\frac{3c^\prime g_sM^2}{2\zeta^{1/3}\mathcal{V}^{4/3}} & \frac{c^\prime g_sM^2}{\zeta^{4/3}\mathcal{V}^{2/3}}
\end{pmatrix},
\end{equation}
up to terms subleading in $1/\mathcal{V}$. Expanding around a vacuum, $\phi^i=\phi^i_0+\delta\phi^i$, we have $\mathcal{L} = -(Q_0)_{ij} (\partial_\mu\delta\phi^i)(\partial^\mu\delta\phi^j) - \frac{1}{2}(M_0)_{ij}\delta\phi^i \delta\phi^j + \ldots$, where $M_{ij}=\frac{\partial^2 V}{\partial \phi^i \partial \phi^j}$ and the subscript $0$ denotes evaluation at the vacuum.
We now perform a basis change $\delta\varphi^i = (\Omega_0^{-1})^i{}_j\delta\phi^j$ with
\begin{equation}
\Omega_0=\begin{pmatrix}
\frac{2\tau_s^{1/4}\sqrt{\mathcal{V}}}{\sqrt{3}\sqrt{\kappa_s}} & 0 & 0 \\
0 & \frac{\sqrt{2}\mathcal{V}^{2/3}}{\sqrt{3}} & 0 \\
0 & 0 & \frac{\zeta^{2/3}\mathcal{V}^{1/3}}{\sqrt{2 c^\prime g_sM^2}}
\end{pmatrix} \label{m0}
\end{equation}
such that $\Omega_0^T Q_0 \Omega_0=\frac{1}{2}\cdot 1_3$ up to terms subleading in $1/\mathcal{V}$.
The mass matrix for the canonically normalized fields is thus
\begin{equation}
\tilde M_0 = \Omega_0^T M_0 \Omega_0. \label{cnmm}
\end{equation}

Using \eqref{m0} and \eqref{lvs-potential}, we can compute $\tilde M_0$ and its eigenvalues explicitly. The characteristic polynomial is $-\lambda^3+b\lambda^2+c\lambda+d$, where, up to small corrections, $b=A+B$, $c=-(A+C)B$, $d=ABC$ with
\begin{align}
A &= \frac{3}{32\pi^2 c^\prime} \sqrt{9-\frac{64\pi c^\prime c^{\prime\prime}}{g_sM^2}} \frac{4\zeta^{2/3}}{9c^\prime g_sM^2 \mathcal{V}^{2/3}},
\qquad B = \frac{2g_s|W_0|^2a_s^2\tau_s^2}{\mathcal{V}^{2}}, \notag \\ C &= 
\frac{9g_s|W_0|^2\kappa_s\sqrt{\tau_s}}{4a_s \mathcal{V}^{3}} \left(\frac{9}{4}- \alpha + \mathcal{O}(g_s)\right). \label{abc}
\end{align}
At large volumes, we have $B\gg C$. One furthermore finds that, generically, $B \gg A \gg C$.\footnote{One might naively think that, for $\mathcal{V}\gg 1$, $B$ should always be smaller than $A$  due to its stronger volume suppression. However, this is not necessarily true as the prefactor of $B$ is much larger than that of $A$.} However, other regimes are possible. For example, for $g_sM^2$ sufficiently close to its lower bound \eqref{gsm}, $A$ can be much smaller than $C$.
The eigenvalues of $\tilde M_0$ are the roots of the characteristic polynomial. Using $B\gg C$, we find
\begin{equation}
m_1^2= A, \qquad m_2^2 = B, \qquad m_3^2 = C \label{m}
\end{equation}
up to small corrections.

An interesting observation is that the mass matrix exhibits significant mass mixing so that the eigenvalues $m_i^2$ are not identical to the moduli masses $m_\zeta^2$, $m_{\tau_s}^2$,  $m_{\tau_b}^2$. One can check that $m_1^2=m_\zeta^2$ and $m_2^2=m_{\tau_s}^2$ up to subleading corrections. However, $m_3^2$ is parametrically lighter than $m_{\tau_b}^2$. In particular, using $\tau_s\sim \mathcal{O}(1)/g_s$, one finds $m_3^2\sim \sqrt{g_s}|W_0|^2/\mathcal{V}^3$, which differs from $m_{\tau_b}^2\sim |W_0|^2/\sqrt{g_s}\mathcal{V}^3$
by a factor $g_s$. As we will discuss in Section \ref{npns}, this is again explained by the NPNS.

Returning to the question of stability, we find from \eqref{abc}, \eqref{m} that, at small $g_s$, $m_3^2$ is positive for $\alpha<\frac{9}{4}$. Together with the earlier requirement $\alpha>1$, we thus find that dS minima lie in the range
\begin{equation}
\alpha \in \Big]1,\frac{9}{4}\Big[. \label{range}
\end{equation}
Note that this interval is wider than the interval $\alpha \in ]1,\frac{5}{4}[$ reported in \cite{Crino:2020qwk}. This is because our \eqref{abc}, \eqref{m} differ from the corresponding expressions in \cite{Crino:2020qwk}. We confirmed our analytic results by comparing them to numerical calculations, finding very good agreement.

\section{Corrections}
\label{vol}

In this section, we discuss various corrections to the potential \eqref{lvs-potential} which can affect the LVS minimum. We will not attempt to be complete but rather discuss a selection of corrections that are potentially relevant for our purpose. See also \cite{Cicoli:2021rub} for a recent survey of perturbative corrections in IIB/F-theory compactifications.

\subsection{Curvature/Loop Corrections to the K\"ahler Potential}
\label{corr1}

\subsubsection{$\alpha^{\prime 2}$ Corrections}
\label{loop}

It is well known that the K\"ahler potential in toroidal orientifold compactifications receives string-loop corrections at order $\alpha^{\prime 2}$ due to an exchange of Kaluza-Klein modes between D7/D3 branes or O7/O3 planes \cite{Berg:2005ja, Berg:2014ama, Haack:2015pbv, Haack:2018ufg} (see also \cite{Garcia-Etxebarria:2012bio}). The general form of these corrections in the CY case is expected to be \cite{Berg:2007wt, Cicoli:2007xp}
\begin{equation}
\delta \mathcal{K} = \sum_i\frac{g_s\sqrt{\tau_i}\,\mathcal{C}_i^\text{KK}(\zeta)}{\mathcal{V}}, \label{dk}
\end{equation}
where $\mathcal{C}_i^\text{KK}$ are unknown functions of the complex-structure moduli.\footnote{Here we focus on a dependence on $\zeta=|Z|$ for simplicity and do not consider a possible dependence on the phase of $Z$.} We will assume $\mathcal{C}_i^\text{KK}$, $\partial_\zeta \mathcal{C}_i^\text{KK}$, $\partial^2_\zeta \mathcal{C}_i^\text{KK} \lesssim \mathcal{O}(1)$ in the following.
In general, there can also be corrections due to winding strings \cite{Berg:2005ja}.
However, these corrections are absent when the two divisors do not intersect (as in the case relevant for us) \cite{Berg:2007wt}.

Due to its specific scaling with respect to the K\"ahler moduli, \eqref{dk} satisfies an extended no-scale structure \cite{vonGersdorff:2005bf, Berg:2005yu, Berg:2007wt, Cicoli:2007xp}.
This means that \eqref{dk} does not give corrections to the LVS potential which are leading in the volume expansion, even though \eqref{dk} is less volume-suppressed than the $\xi$ term in \eqref{k0}.
Computing the $F$-term scalar potential including the correction \eqref{dk}, we find that the leading corrections to \eqref{lvs-potential} are
\begin{align}
\delta V &= \frac{2a_s^2|A_s|^2g_s^2 \mathcal{C}_s^\text{KK}\e^{-2a_s\tau_s}}{9\kappa_s^2 \sqrt{\tau_s} \mathcal{V}} - \frac{2a_s|A_s|g_s^2\mathcal{C}_s^\text{KK}|W_0|\e^{-a_s\tau_s}}{3\kappa_s\mathcal{V}^2} + \frac{g_s^3(\mathcal{C}_s^\text{KK})^2|W_0|^2}{12\kappa_s\sqrt{\tau_s}\mathcal{V}^3} \nl + \frac{3c^\prime g_s^3 M^2 \zeta^{2/3} \mathcal{C}^\text{KK}_b|W_0|^2}{\mathcal{V}^{10/3}}. \label{dv}
\end{align}
All other terms, including those involving $\partial_\zeta \mathcal{C}_i^\text{KK}$ or $\partial^2_\zeta \mathcal{C}_i^\text{KK}$,
can be shown to be
negligible in the relevant regime (under the above assumption of $\lesssim\mathcal{O}(1)$ coefficients).
Note that the first line of \eqref{dv} agrees with an earlier result in \cite{Berg:2007wt} that did not consider the $Z$ and $Y$ fields in $\mathcal{K}$ and $W$.
The term in the second line is new and arises when the light conifold modulus is taken into account. On the other hand, the presence of the anti-brane uplift (i.e., the field $Y$) does not affect how \eqref{dk} corrects the scalar potential.\footnote{It is conceivable that the presence of the anti-brane results in further loop corrections beyond the term \eqref{dk} arising in $\mathcal{N}=1$ vacua. However, we are not aware of any concrete results on this in the literature.}

Using $\tau_s=\mathcal{O}(1)/g_s$, we find that the first two terms in \eqref{dv} are suppressed by a factor $g_s^2$ compared to the first two terms in \eqref{lvs-potential}, while the third term is suppressed by a factor $g_s^4$ compared to the third term in \eqref{lvs-potential}. The first two terms in \eqref{dv} are therefore more relevant.\footnote{This is consistent with the results of \cite{Berg:2007wt} but disagrees with \cite{Cicoli:2007xp}, where it was claimed that the loop corrections to the non-perturbative part of the scalar potential are negligible compared to the $(\mathcal{C}_s^\text{KK})^2$ term.}
The term in the second line of \eqref{dv} is naively volume-suppressed compared to those in the first line, but only by a small power $\mathcal{V}^{-1/3}$ and with a potentially large coefficient.\footnote{The last term in \eqref{dv} might compete with terms generated by $\zeta$-dependent warping corrections to the ansatz \eqref{dk}, which is only expected to be valid in the unwarped approximation.
Similarly, one may wonder whether the $(\mathcal{C}^\text{KK}_s)^2$ term in \eqref{dv} could compete with higher-loop corrections.
However, we are not aware of any argument that such possible competing terms would imply cancellations in \eqref{dv}.} It can therefore be the dominant correction, as we checked in explicit solutions in the model of Section \ref{dsmin}.

Finally, let us note that there are also $\alpha^{\prime 2}$ corrections already at tree-level in $g_s$.
These corrections are absorbed into a redefinition of the K\"ahler moduli \cite{Grimm:2013bha, Grimm:2014efa, Grimm:2015mua}, i.e., $\mathcal{K}$ is not corrected once it is expressed in terms of the properly defined $T_i$. Alternatively, these corrections can be understood as artifacts of an inconvenient field frame in M-theory \cite{Junghans:2014zla}. As explained in \cite{Junghans:2014zla}, they have no physical meaning but are mere ``gauge'' choices.

\subsubsection{Log field redefinitions}
\label{sec:log}

Another type of corrections arises from one-loop field redefinitions of the K\"ahler moduli of the form $\tau_i^\text{new} = \tau_i^\text{old} + \mathcal{C}^\text{log}_i \ln\mathcal{V}$ with (in general unknown) coefficients $\mathcal{C}^\text{log}_i$. The field redefinitions are related to threshold corrections to gauge couplings and imply a correction to the K\"ahler potential after expressing the physical CY volume in terms of the corrected moduli $\tau_i^\text{new}$.
The evidence for the redefinitions comes from orbifold calculations \cite{Conlon:2009xf, Conlon:2009kt}, field-theory arguments \cite{Conlon:2010ji} and dimensional reduction \cite{Grimm:2017pid, Weissenbacher:2019mef, Weissenbacher:2020cyf}.
Furthermore, it was argued that such terms are in some cases required for the consistency of heterotic/F-theory duality in certain infinite-distance limits \cite{Klaewer:2020lfg}.

It is instructive to briefly review the argument of \cite{Conlon:2010ji}. Consider a stack of D-branes wrapping a 4-cycle with volume $\tau^\text{old}$. The running gauge coupling of the worldvolume gauge theory is then
\begin{equation}
\frac{1}{g^2(\mu)} = \frac{\tau^\text{old}}{4\pi}+\frac{\beta}{16\pi^2}\ln\left( \frac{\Lambda_\text{UV}^2}{\mu^2} \right),
\end{equation}
where $\mu$ is the energy scale, $\Lambda_\text{UV}$ is the UV cutoff and $\beta$ is related to the beta function. The theory thus becomes strongly coupled at the scale $\Lambda_\text{strong}\sim \Lambda_\text{UV} \e^{\frac{2\pi\tau^\text{old}}{\beta}}$. Assuming that gaugino condensation generates a non-perturbative superpotential, we furthermore have
\begin{equation}
W \sim \e^{\frac{6\pi T}{\beta}} \label{wt}
\end{equation}
in Planck units, where $T$ is the K\"ahler modulus which classically satisfies $\text{Re}T=\tau^\text{old}$.
Since the superpotential is generated at the strong-coupling scale, we further assume that the corresponding K\"ahler-invariant quantity satisfies
\begin{equation}
\e^{\mathcal{K}/2}|W| \sim \Lambda_\text{strong}^3 \sim \Lambda_\text{UV}^3 \e^{\frac{6\pi\tau^\text{old}}{\beta}}.
\end{equation}
Substituting \eqref{wt} and solving for $\text{Re}T=\tau^\text{new}$, we thus find
\begin{align}
\tau^\text{new} &= \tau^\text{old} - \frac{\beta \mathcal{K}}{12\pi} + \frac{\beta}{2\pi}\ln \Lambda_\text{UV} \label{tt}
= \tau^\text{old} - \frac{\beta}{12\pi}\ln\mathcal{V},
\end{align}
where we used $\mathcal{K}=-2\ln\mathcal{V}+\ldots $ and identified $\Lambda_\text{UV}$ with the string scale (i.e., $\Lambda_\text{UV}\sim \mathcal{V}^{-1/2}$ in Planck units) in the second step.\footnote{Here we focus on the (leading) volume scaling of the correction and ignore, e.g., other terms in $\mathcal{K}$, which could give subleading corrections to the right-hand side of \eqref{tt}.} We thus reproduce under these assumptions the above claim that the definition of the K\"ahler modulus is shifted at the one-loop level by a logarithmic term.

In the context of the LVS, the best motivated case is a redefinition of $\tau_s$ \cite{Conlon:2010ji}. We will therefore assume that \eqref{ve} is replaced by $\mathcal{V} = \tau_b^{3/2} - \kappa_s \left(\tau_s- \mathcal{C}^\text{log}_s \ln \mathcal{V}\right)^{3/2}$, where $\tau_i=\text{Re}(T_i)$ are the redefined K\"ahler moduli and we consider constant $\mathcal{C}^\text{log}_s$ as in \cite{Conlon:2010ji}. At linear order in $\mathcal{C}^\text{log}_s$, we find that the leading corrections to the LVS potential \eqref{lvs-potential} are
\begin{align}
\delta V &= - \frac{\mathcal{C}^\text{log}_s \ln\mathcal{V}}{\tau_s}\frac{2 a_s^2|A_s|^2g_s\sqrt{\tau_s}\e^{-2a_s\tau_s}}{3\kappa_s \mathcal{V}} + \frac{\mathcal{C}^\text{log}_s \ln\mathcal{V}}{\tau_s} \frac{2 a_s|A_s|g_s\tau_s |W_0|\e^{-a_s\tau_s}}{\mathcal{V}^2} \nl - \frac{9 g_s \kappa_s\sqrt{\tau_s} \,\mathcal{C}^\text{log}_s |W_0|^2}{4 \mathcal{V}^3}. \label{log}
\end{align}
This agrees with an earlier result of \cite{Conlon:2010ji}\footnote{Except for a disagreement on a phase factor in the second term of \eqref{log}.}, where a two-field model without the $Z$ and $Y$ fields was considered. Indeed, all $\mathcal{C}^{\text{log}}_s$ corrections to the scalar potential that depend on the conifold modulus turn out to be subleading compared to \eqref{log}. One can also check that no corrections to the anti-brane uplift (i.e., no terms $\sim c^{\prime\prime}$) are generated in the scalar potential by the field redefinition.

Note that the last term in \eqref{log} is suppressed by a factor $g_s^{3/2}\sqrt{\tau_s}\sim g_s$ relative to \eqref{lvs-potential} (using that $\tau_s=\mathcal{O}(1)/g_s$).
On the other hand, the first two terms in \eqref{log} scale like $\ln\mathcal{V}/\tau_s$ relative to \eqref{lvs-potential}, where $\ln\mathcal{V}\sim \tau_s$ on-shell.
We conclude that these terms are not suppressed by $g_s$ (or any other small parameter).

\subsubsection{$\alpha^{\prime 3}$ Corrections}
\label{loop2}

Next, we consider $\alpha^{\prime 3}$ corrections to the K\"ahler potential. One such correction was computed in \cite{Minasian:2015bxa}. It shifts the BBHL \cite{Becker:2002nn} term $\sim \xi$ in \eqref{k0} as follows:
\begin{equation}
\mathcal{K} = -2 \ln\left(\mathcal{V} + \frac{\xi-\Delta\xi}{2g_s^{3/2}} \right) + \ldots \label{kp}
\end{equation}
The shift $\Delta\xi$ is given by
\begin{equation}
\Delta\xi = \frac{\zeta(3)}{(2\pi)^3}\int_X D_\text{O7}^3,
\end{equation}
where
$D_\text{O7}$ is the Poincar\'e dual of the divisor wrapped by the O7 planes.\footnote{This result was obtained in \cite{Minasian:2015bxa} in the weak-coupling limit of a smooth 4-fold. However, the authors argued that it should hold in singular 4-folds as well (although possibly supplemented by additional terms). It was furthermore pointed out in \cite{Minasian:2015bxa} that a complete dimensional reduction of the kinetic terms of the K\"ahler moduli might reveal further corrections at the same order in $\alpha^\prime$ and $g_s$.}

The K\"ahler potential also receives $g_s$ corrections at the same order in $\alpha^\prime$.
Such corrections are expected to appear at open/unoriented string tree-level ($\sim g_s$) \cite{Minasian:2015bxa, Antoniadis:2019rkh}, one loop ($\sim g_s^2$) \cite{Antoniadis:1997eg, Berg:2014ama, Haack:2015pbv, Haack:2018ufg, Minasian:2015bxa} and higher orders in $g_s$ \cite{Minasian:2015bxa, Antoniadis:2019rkh}.
The terms linear (and cubic) in $g_s$ can acquire an $\ln\mathcal{V}$ dependence due to an exchange of KK modes between 10D $R^4$ terms and distant D7 branes/O7 planes \cite{Antoniadis:2019rkh}. Unfortunately, the coefficients of the various corrections are not known for general CY orientifolds.

The corrections can be taken into account by making the replacement
\begin{equation}
\xi \to \xi -\Delta\xi + \mathcal{C}^{\xi}_1\, g_s \ln\mathcal{V} + \mathcal{C}^{\xi}_2 g_s +\ldots \label{rep}
\end{equation}
in \eqref{k0} with some coefficients $\mathcal{C}^{\xi}_i$, which we will assume to be $\lesssim \mathcal{O}(1)$ constants for simplicity.
The $\Delta\xi$ and $\mathcal{C}^{\xi}_2$ terms in \eqref{rep} only shift $\xi$ by a constant. The correction to the LVS potential due to these terms is thus simply a shift in the coefficient of the third term in \eqref{lvs-potential}. One can furthermore verify that the leading correction to the LVS potential at linear order in $\mathcal{C}^{\xi}_1$ is also captured by such a shift. The correction to the LVS potential is thus
\begin{equation}
\delta V = \frac{3|W_0|^2 \left(-\Delta\xi + \mathcal{C}^{\xi}_1\, g_s \ln\mathcal{V} + \mathcal{C}^{\xi}_2 g_s\right)}{8\sqrt{g_s}\mathcal{V}^3}. \label{dv2}
\end{equation}
Note that the term $\mathcal{C}^{\xi}_1$ is not suppressed by any small parameter relative to the leading LVS potential (because of $\ln\mathcal{V} \sim \tau_s \sim 1/g_s$), similarly to the field-redefinition corrections discussed in Section \ref{sec:log}.

\subsection{Curvature Corrections to the Gauge-Kinetic Function}
\label{corr2}

Another correction we will consider enters the gauge-kinetic function of a D7 brane \cite{Jockers:2004yj, Jockers:2005zy, Haack:2006cy} and thus affects the non-perturbative superpotential generated by gaugino condensation. The correction can be derived considering the well-known curvature corrections to the D-brane action \cite{Bachas:1999um, Bershadsky:1995qy, Green:1996dd, Cheung:1997az, Minasian:1997mm}.
Since we are interested in corrections that modify the gauge-kinetic function, we focus on terms $\sim\mathcal{F}_{\mu\nu}^2$, where $\mathcal{F}=B_2+2\pi\alpha^\prime F_2$ and $F_2$ is the worldvolume-gauge-field strength. For simplicity, we consider compactifications with $h^{1,1}_-(X)=0$ such that we can set $C_2=0$. The relevant terms in the CS action are then (in units where $2\pi\sqrt{\alpha^\prime}=1$)
\begin{align}
S_\text{D7} \supset \pi \int \mathcal{F} \w \mathcal{F} \int_{\Sigma_i} &\left[ C_4 + \frac{1}{2}\mathcal{F} \w \mathcal{F} C_0 + \frac{p_1(T\Sigma_i)-p_1(N\Sigma_i)}{48} C_0 \right]. \label{sd7}
\end{align}
Here $\Sigma_i$ is the 4-cycle wrapped by the D7 brane and $p_1(T\Sigma_i)$, $p_1(N\Sigma_i)$ are the first Pontryagin classes of the tangent/normal bundle.
On a CY, the adjunction formula implies $\int_{\Sigma_i}(p_1(T\Sigma_i)-p_1(N\Sigma_i))/48=-\chi_i/24=Q_3^{\text{D7}}$ \cite{Cheung:1997az}. Here, $\chi_i\equiv\chi(\Sigma_i)$ is the Euler number and $Q_3^{\text{D7}}$ is the induced D3-brane charge on the D7 brane.
In the weak-coupling limit of F-theory compactifications, D7 branes often wrap singular surfaces such that this formula needs to be generalized appropriately 
\cite{Aluffi:2007sx, Braun:2008ua, Collinucci:2008pf}.

We can now read off the gauge-kinetic function $f_i$ from the factor multiplying $\mathcal{F} \w \mathcal{F}$ in \eqref{sd7}:
\begin{equation}
\text{Im} (f_i)=\text{Im}(T_i) + \left(\frac{1}{2}\int_{\Sigma_i} \mathcal{F} \w \mathcal{F} -\frac{\chi_i}{24}\right) \text{Im}(S),
\end{equation}
where $T_i$ is the K\"ahler modulus associated to the 4-cycle $\Sigma_i$ and $S$ is the axio-dilaton.
Analogous curvature corrections in the DBI action holomorphically complete this to\footnote{One may wonder whether the $S$ dependence in \eqref{gkf} is in conflict with the non-renormalization theorems of \cite{Witten:1985bz, Dine:1986vd, Burgess:2005jx, GarciadelMoral:2017vnz}. In particular, it was argued in \cite{Burgess:2005jx} that $f_i$ cannot receive $S$-dependent corrections to all orders in perturbation theory. However, this result was obtained under the assumption of a shift symmetry of $S$, which is broken by the CS terms \eqref{sd7}. The results of \cite{Burgess:2005jx} are therefore not in conflict with \eqref{gkf}.}
\begin{equation}
f_i=T_i+\left(\frac{1}{2}\int_{\Sigma_i} \mathcal{F} \w \mathcal{F} -\frac{\chi_i}{24}\right) S. \label{gkf}
\end{equation}

For simplicity, we now focus on a stack of D7 branes with vanishing flux $\mathcal{F}$ on $\Sigma_i$. Gaugino condensation on such a stack can generate a non-perturbative superpotential $W_\text{np}\sim \e^{-a_if_i}$ \cite{Kachru:2003aw}. We thus find
\begin{equation}
W_\text{np}= A_i \e^{-a_i(T_i-\chi_i S/24)}. \label{wnp}
\end{equation}
The same form of a non-perturbative superpotential is obtained from a Euclidean D3-brane instanton wrapping $\Sigma_i$ \cite{Denef:2005mm}. This follows because the CS action of a Euclidean D3 brane has exactly the same structure as \eqref{sd7}:
\begin{equation}
S_\text{E3} \supset -2\pi i \int_{\Sigma_i} \left[ C_4 + \frac{1}{2}\mathcal{F} \w \mathcal{F} C_0 + \frac{p_1(T\Sigma_i)-p_1(N\Sigma_i)}{48} C_0 \right].
\end{equation}
If the instanton has the right number of zero modes, it generates a non-perturbative superpotential $W_\text{np}\sim \e^{-S_\text{E3}}$ \cite{Witten:1996bn}. Assuming again an absence of worldvolume flux for simplicity, we obtain the same form of the superpotential as in \eqref{wnp}.

In the context of the LVS, \eqref{wnp} amounts to the replacement
\begin{equation}
|A_s|\to |A_s| \e^{a_s\chi_s/24g_s}
\end{equation}
in the LVS potential. Note that, in practice, $A_s$ is often not known explicitly but set to a value such as $A_s=1$ by hand. Keeping track of the curvature correction is therefore in some sense arbitrary, as we could just as well absorb it into the definition of $A_s$.
Nevertheless, the ultimate goal should be to construct flux vacua that are as explicit as possible, including a computation of constants such as $A_s$.
We will therefore keep the correction explicit and check how it affects the LVS minimum at a fixed value of $A_s$.

\subsection{Curvature/Warping Corrections from Conifold-Flux Backreaction}
\label{corr3}

We now move on to another type of correction, which was recently argued in \cite{Carta:2019rhx, Gao:2020xqh} to arise due to large fluxes at the tip of the conifold region.

As is well known, both warping and $\alpha^\prime$ corrections become negligible in the large-volume limit, i.e., taking $\mathcal{V}_s\to\infty$ (with $\mathcal{V}_s$ the string-frame volume) while keeping other quantities such as the flux numbers fixed.
A stronger assumption often used in the literature is that a sufficient condition to neglect the corrections is $\mathcal{V}_s\gg 1$ (in string units). However, this is in general not true. In particular, charged objects such as branes, O-planes and fluxes generate field gradients proportional to their charges. If these charges are sufficiently large, the warp factor varies strongly and the 10D curvature invariants/energy densities can be large even at volumes much larger than 1.

In particular, it was shown in \cite{Carta:2019rhx} that KKLT dS models \cite{Kachru:2003aw} with $h^{1,1}=1$ are generically incompatible with a weakly-warped CY bulk. The problem occurs because the KKLT scenario requires a large charge $KM$ dissolved in fluxes in the conifold region, while the volume at which the dS vacuum is stabilized is comparatively small (even though still larger than 1 in string units).
Even worse, \cite{Gao:2020xqh} showed that, for any $h^{1,1}$, the KKLT scenario generically suffers from a ``singular-bulk problem'', i.e., the field gradients become so large that the curvature diverges and the metric is formally singular in large parts of the CY. 10D supergravity is then not a reliable description of such a compactification anymore.\footnote{This should be contrasted with the usual O-plane singularities, which are localized within a sub-stringy diameter around the O-planes. Although the 10D supergravity description breaks down in that case as well, at large volume this only happens in a parametrically small fraction of the compactification space and is thus expected to be negligible from the point of view of the 4D EFT. On the other hand, in KKLT dS vacua, the singularities surrounding the O-planes grow to a much larger size.} Note that the main issue here is not the singularity itself---indeed, it is plausible that it is resolved by string theory in some way (see \cite{Carta:2021lqg} for an interesting proposal). The problem is rather that the 4D EFT describing this highly stringy regime is unknown and might look very different from the naive EFT derived from a dimensional reduction of 10D supergravity.\footnote{It was argued in \cite{Carta:2021lqg} that the KKLT dS minimum survives beyond the supergravity regime. However, this is not convincing, as the K\"ahler potential is then an unknown function due to large curvature and warping corrections. A further problem is that, according to \cite{Carta:2021lqg}, the resolution of the singularity leads to a recombination of positive and negative D3-brane charges in the bulk. This is worrisome, as it may significantly reduce the tadpole compared to the naive one and thus limit the fluxes that can be used for the stabilization of the complex-structure moduli.}

It follows from \cite{Carta:2019rhx, Gao:2020xqh} that a necessary condition to avoid the singular-bulk problem is
\begin{equation}
\frac{\mathcal{V}^{2/3}}{KM} \gg 1. \label{lv1}
\end{equation}
The same condition ensures weak warping in the bulk.
On the other hand, for $\mathcal{V}^{2/3}\lesssim KM$, we enter a regime of large warping and large curvature, and a variety of corrections is expected to blow up.

Perhaps surprisingly, the singular-bulk problem is also relevant for the LVS. As already noted in \cite{Gao:2020xqh}, LVS models can in principle avoid the problem as they can have an extremely large volume. However, we will see below that this is not always true. In particular, one can show that, in regions of the parameter space where the singular-bulk problem is absent, either other types of corrections blow up or there are no dS vacua. Conversely, for those parameter choices that yield dS vacua and have all other corrections suppressed, one always finds a singular-bulk problem.

We will not attempt to analyze all possible corrections that blow up when \eqref{lv1} is violated but only give an example here.
Recall from Section \ref{loop2} that the BBHL term $\sim\xi$ in the K\"ahler potential is expected to receive a $g_s$ correction.
The existence of such a correction is suggested by a correction to the 4D Einstein-Hilbert term that arises from dimensionally reducing an $R^4$ term with a varying dilaton \cite{Minasian:2015bxa}. We now argue that an analogous correction to the BBHL term should arise for a varying warp factor. Indeed, one can verify that dimensionally reducing the 10D $R^4$ terms of type IIB string theory yields a correction to the 4D Einstein-Hilbert term proportional to\footnote{The F-theory generalization of this expression can be obtained following \cite{Minasian:2015bxa}.}
\begin{equation}
\int_X \e^{2A(y)-3\phi(y)/2} c_3(X) = g_s^{-3/2}\left[ \chi(X) + \mathcal{O}\left(g_s\right) + \mathcal{O}\left(\frac{KM}{\mathcal{V}^{2/3}}\right) \right]. \label{cflux0}
\end{equation}
Here
$\e^{2A(y)}$ is the Einstein-frame warp factor, $\phi(y)$ is the dilaton, $c_3(X)$ is the third Chern class and $\chi(X)$ is the Euler characteristic.
The right-hand side of \eqref{cflux0} is valid at large volume and small $g_s$, where we split the dilaton and the warp factor into a constant and a varying part such that $\e^{-\phi(y)}=1/g_s+\e^{-\phi_0(y)}$ and $\e^{-4A(y)} = 1 + \e^{-4A_0(y)}/\mathcal{V}^{2/3}$ with $\e^{-4A_0(y)}=\mathcal{O}(KM)$ in the bulk \cite{Giddings:2005ff, Gao:2020xqh}. Note that, aside from \eqref{cflux0}, various other 10D higher-derivative terms may in principle yield further corrections to the 4D Einstein-Hilbert term, e.g., involving warp-factor and dilaton derivatives or powers of $F_5$ and $G_3$.\footnote{For example, expressing $R^4$ in terms of the unwarped metric produces terms involving warp-factor derivatives
in addition to the terms in \eqref{cflux0}. In particular, terms containing a factor $\nabla_m\partial_n\e^{-4A}$ and two internal Riemann tensors may appear at the same order in the volume expansion as the warping correction in \eqref{cflux0}, while all other warping terms in $R^4$ are suppressed by higher powers of the volume. This can be verified using $\e^{-4A} = 1 + \e^{-4A_0}/\mathcal{V}^{2/3}$ and the metric of \cite{Giddings:2001yu} in the expressions in App.~B of \cite{Junghans:2013xza}. By similar arguments, one finds that corrections to the Einstein-Hilbert term from $F_5^2R^3$, $F_5^4R^2$ or $F_5^6R$ terms are volume-suppressed compared to \eqref{cflux0} as well.}
A complete analysis of all such corrections is beyond the scope of this work. However, we do not expect a cancellation effect due to such terms, as they have a different dependence on the 10D fields and are not related by the equations of motion to \eqref{cflux0}.

We now observe that the first term on the right-hand side of \eqref{cflux0} is related to the usual BBHL term in \eqref{k0} \cite{Becker:2002nn, Bonetti:2016dqh}, while the second term indicates the presence of the $g_s$ correction due to the varying dilaton. Crucially, the third term suggests a further correction due to the varying warp factor, i.e., a shift in \eqref{k0} of the form
\begin{equation}
\xi \to \xi + \frac{\mathcal{C}^\text{flux} KM}{\mathcal{V}^{2/3}}. \label{cflux}
\end{equation}
Here $\mathcal{C}^\text{flux}$ is an unknown function of the moduli, which, for concreteness, we will assume to be a constant $\lesssim \mathcal{O}(1)$.
In order to derive $\mathcal{C}^\text{flux}$, we would have to determine the $\alpha^\prime$-corrected kinetic terms of the various moduli by dimensional reduction (see, e.g., \cite{Grimm:2014efa, Grimm:2015mua, Berg:2014ama, Haack:2015pbv, Haack:2018ufg, Bonetti:2016dqh}) and explicitly compute the warp factor (i.e., the Green's function of the Laplacian on the CY \cite{Giddings:2001yu}), which is beyond the scope of this work.

The leading correction to the LVS potential \eqref{lvs-potential} due to \eqref{cflux} is
\begin{equation}
\delta V =  \frac{15\mathcal{C}^\text{flux} KM |W_0|^2}{8\sqrt{g_s}\mathcal{V}^{11/3}}. \label{cflux2}
\end{equation}
This is naively volume-suppressed compared to the terms in \eqref{lvs-potential}. However, due to the factor $KM$, the coefficient can be quite large such that the correction is still relevant.

\subsection{Curvature Corrections in the Conifold Region}
\label{corr4}

A further type of potentially relevant corrections are curvature corrections at the tip of the deformed-conifold region. Such corrections may affect the anti-brane uplift and the stabilization of the conifold modulus. In particular, the string-frame radius of the $S^3$ at the conifold tip satisfies $R_{S^3}^2\sim g_sM \alpha^\prime$ \cite{Klebanov:2000hb, Herzog:2001xk}. We therefore expect that the part of the LVS potential describing the conifold region (i.e., the second line of \eqref{lvs-potential}) receives corrections suppressed by powers of $1/g_sM$.

Performing a dimensional reduction of all relevant $\alpha^\prime$ corrections is beyond the scope of this paper.
However, as an explicit example, recall that brane actions receive $R^2$ curvature corrections \cite{Bachas:1999um, Bershadsky:1995qy, Green:1996dd, Cheung:1997az, Minasian:1997mm}.
In particular, the $R^2$ corrections to the DBI action of the anti-D3 brane involve components of the Riemann tensor with two tangent and two normal indices \cite{Bachas:1999um}, which are non-zero in the warped deformed conifold.
Explicitly, we find (in the string frame)
\begin{align}
\mathcal{L}_\text{DBI} &\supset - \mu_3 \e^{-\phi} \left[ 1 - \frac{(4\pi^2\alpha^\prime)^2}{12\cdot 32\pi^2} R_{a \alpha b}{}^\alpha R^{a}{}_\beta{}^{b\beta} \right] = - \mu_3 \e^{-\phi} \left[ 1 - 16\frac{(4\pi^2\alpha^\prime)^2}{12\cdot 32\pi^2} |\nabla_a\partial_b A|^2 \right] \nll =
- \mu_3 \e^{-\phi} \left[ 1 - \frac{6^{2/3}(4\pi^2)^2}{3\cdot 32\pi^2 (g_sM)^2} \frac{I^{\prime\prime}(0)^2}{I(0)^3} \right] = -\mu_3 \e^{-\phi}\left[ 1 - \frac{1.97}{(g_sM)^2} \right]. \label{itau}
\end{align}
Here, $\alpha$, $\beta$ are tangent indices, $a$, $b$ are normal indices, $\e^{2A}$ is the warp factor as before
and $I(\tau)$ is a function defined in \cite{Herzog:2001xk}. We also used that the first derivative of the warp factor vanishes at the conifold tip and substituted the explicit expressions for the warp factor and the 6D metric from \cite{Klebanov:2000hb, Herzog:2001xk}.

Note that terms with the same scaling as in \eqref{itau} should also arise, e.g., from 4-derivative corrections to the DBI action involving powers of the 3-form field strength $G_3$. The existence of such terms can be inferred from T-duality arguments (see \cite{Robbins:2014ara} for a derivation in the case of O-plane actions), and their scaling with $g_sM$ follows from the expressions for $G_3$ and the metric provided in \cite{Klebanov:2000hb, Herzog:2001xk}.

We conclude that the tension of the anti-D3 brane (and therefore the uplift term $\sim c^{\prime\prime}$ in the LVS potential \eqref{lvs-potential}) receives corrections at the order $1/(g_sM)^2$. Barring an unknown cancellation effect, we thus have
\begin{equation}
\delta V = \frac{\mathcal{C}^\text{con}}{(g_sM)^2} \frac{c^{\prime\prime}\zeta^{4/3}}{2\pi g_sM^2\mathcal{V}^{4/3}}. \label{ccon}
\end{equation}
Here $\mathcal{C}^\text{con}$ is a function of the moduli, which, for concreteness, we will assume to be constant and $\lesssim\mathcal{O}(1)$.

\subsection{Higher $F$ terms}
\label{corr5}

Finally, the scalar potential receives corrections from higher $F$ terms, which arise from terms with 4 superspace derivatives in the 4D EFT \cite{deAlwis:2012vp, Cicoli:2013swa, Ciupke:2015msa}.
It was argued in \cite{Cicoli:2013swa} that such corrections are generated by integrating out KK modes and that their suppression is controlled by the parameter
\begin{equation}
\frac{m_{3/2}^2}{m_\text{KK}^2} \sim \frac{g_s|W_0|^2}{\mathcal{V}^{2/3}}. \label{cf}
\end{equation}
Here $m_{3/2} \sim \sqrt{g_s}|W_0|/\mathcal{V}$ is the gravitino mass and $m_\text{KK} \sim \mathcal{V}^{-2/3}$ is the KK scale (both in Planck units).
Another source of higher $F$ terms are 10D 8-derivative terms containing powers of the 3-form field strength $G_3$.
Dimensionally reducing such terms, one finds that they are suppressed by the same factor \eqref{cf} compared to the $\xi$ term in the LVS potential \eqref{lvs-potential} \cite{Ciupke:2015msa} (see also \cite{Conlon:2005ki, Burgess:2020qsc, Cicoli:2021rub}).

We conclude that the LVS potential receives a correction
\begin{equation}
\delta V = \frac{\mathcal{C}^F g_s|W_0|^2}{\mathcal{V}^{2/3}} \frac{|W_0|^2}{\sqrt{g_s}\mathcal{V}^3}, \label{cf2}
\end{equation}
where $\mathcal{C}^F$ is an unknown, possibly moduli-dependent coefficient. For concreteness, we will assume that $\mathcal{C}^F$ is constant and $\lesssim\mathcal{O}(1)$.

\section{Non-perturbative No-scale Structure}
\label{npns}

Before we discuss how the above corrections affect the LVS vacua, we first need to understand
a special property of the LVS potential which to our knowledge has not been discussed in the literature before.\footnote{The effect of a leading-in-$g_s$ cancellation in the vacuum energy was already observed in \cite{Hebecker:2012aw}. Here we point out an analogous cancellation in the mass matrix and explain the responsible structure of the potential, which crucially has the consequence of enhancing certain types of corrections.} We will call this property the \emph{non-perturbative no-scale structure (NPNS)}, as it is reminiscent of the cancellations in the scalar potential due to the (extended) no-scale structure. The underlying reason for the NPNS is the $1/g_s$ scaling in the exponent of the non-perturbative factor $\e^{-a_s\tau_s}$ (recall that $\tau_s\sim \mathcal{O}(1)/g_s$). As we will now show, this scaling implies that the vacuum energy vanishes in the LVS at leading order in $g_s$, i.e., it is smaller by a factor $g_s$ compared to the individual terms in the off-shell potential. An analogous leading-in-$g_s$ cancellation occurs in one of the eigenvalues of the mass matrix.

To see this, consider for simplicity the leading LVS potential \eqref{lvs-potential} ignoring the $\zeta$ terms and possible corrections. For convenience, we furthermore set $a_s=A_s=\xi=W_0=\kappa_s=1$ for the moment, as these constants do not play a role for our argument. We thus obtain
\begin{align}
V = \frac{4 g_s\sqrt{\tau_s}\e^{-2 \tau_s}}{3\mathcal{V}} - \frac{2 g_s\tau_s \e^{- \tau_s}}{\mathcal{V}^2} + \frac{3}{8\sqrt{g_s}\mathcal{V}^3}.
\end{align}
We now make the field redefinition $\tau_s=t/g_s$, $\mathcal{V}=v\, \e^{t/g_s}/\sqrt{g_s}$, which makes manifest the $g_s$ dependence of the potential near the LVS minimum (cf.~\eqref{tb0}, \eqref{ts0}). We thus obtain a potential which depends on the two fields $v$ and $t$ and the small parameter $g_s$:
\begin{align}
V = g_s \e^{-3 t/g_s} \left( \frac{4\sqrt{t}}{3v} - \frac{2 t}{v^2} + \frac{3}{8v^3} \right). \label{tv}
\end{align}
Near the minimum, we have $t,v\sim\mathcal{O}(g_s^0)$ such that each term in the potential is of the order $\mathcal{O}(g_s\e^{-3 t/g_s})$. Now consider the equation of motion for $t$ at leading order in $g_s$:
\begin{equation}
0=\partial_t V = -3\e^{-3 t/g_s} \left( \frac{4\sqrt{t}}{3v} - \frac{2 t}{v^2} + \frac{3}{8v^3} \right) + \mathcal{O}(g_s\e^{-3 t/g_s}) = -\frac{3}{g_s}V + \mathcal{O}(g_s\e^{-3 t/g_s}).
\end{equation}
It follows from this equation that $V=\mathcal{O}(g_s^2\e^{-3 t/g_s})$ on-shell, which is by a factor $g_s$ smaller than the off-shell potential.

The reason for this cancellation is that, at leading order in $g_s$, only the factor $\e^{-3t/g_s}$ in \eqref{tv} plays a role for the $t$ equation since acting with $\partial_t$ on $\e^{-3t/g_s}$ generates a factor $1/g_s$. On the other hand, derivatives with respect to the $t$-dependent terms in the bracket in \eqref{tv} do not generate such a factor and are therefore subleading in the $t$ equation. The leading potential felt by $t$ is thus of the form $V = \text{const.}\cdot \e^{-3 t/g_s}$, which has no extremum unless the constant factor is zero. At next-to-leading order in $g_s$, the $t$ dependence in the bracket in \eqref{tv} modifies the $t$ equation such that it is no longer proportional to $V$. We then recover the usual LVS AdS minimum.

Analogously, one can show that there is an on-shell cancellation in one of the eigenvalues of the mass matrix at leading order in $g_s$. The eigenvalues can be written in terms of $V$ as
\begin{align}
m_2^2 &= \frac{4v\sqrt{t}\, \e^{t/g_s}}{3 g_s} \left(g_s^2\partial_{t}^2 V-2g_sv\partial_{t}\partial_{v} V+v^2\partial_{v}^2 V\right), \label{m2m3-1} \\ m_3^2 &= \frac{3v^2 g_s^2}{2} \frac{ (\partial_{t}^2 V)(\partial_{v}^2V)- (\partial_{t}\partial_{v}V)^2}{g_s^2\partial_{t}^2 V-2g_sv\partial_{t}\partial_{v} V+v^2\partial_{ v}^2 V}. \label{m2m3-2}
\end{align}
Here we diagonalized the canonically normalized mass matrix as explained in Section \ref{setup} and dropped terms suppressed by $\e^{-t/g_s}$.
Crucially, the NPNS implies an on-shell cancellation in the two components $\partial_{t}^2 V$ and $\partial_{t}\partial_v V$ at leading order in $g_s$.
This follows because $\partial_{t} V = - \frac{3}{g_s} V$ at leading order and therefore $\partial_i \partial_{t} V = - \frac{3}{g_s} \partial_i V = 0$.
We thus find
\begin{align}
g_s^2\partial_{t}^2 V \sim g_s^2 \e^{-3t/g_s}, \qquad g_s\partial_{t}\partial_{v} V \sim g_s^2 \e^{-3t/g_s}, \qquad \partial_{v}^2 V \sim g_s \e^{-3t/g_s} \label{ppv}
\end{align}
on-shell, as can be verified using \eqref{tv}. Note that the first two terms are subleading compared to the last one for small $g_s$. Using this in \eqref{m2m3-1} and \eqref{m2m3-2}, the eigenvalues simplify to
\begin{align}
m_2^2 \sim \frac{\e^{t/g_s}}{g_s} \partial_{v}^2 V \sim \e^{-2t/g_s}, \qquad m_3^2 \sim g_s^2 \partial_{t}^2 V \sim g_s^2 \e^{-3t/g_s}
\end{align}
up to small corrections. We stress again that the scaling of $m_3^2$ relies on the fact that several $\mathcal{O}(g_s \e^{-3t/g_s})$ terms cancel in $g_s^2 \partial_{t}^2 V$ and $g_s\partial_{t}\partial_{v} V$ upon imposing the equations of motion.

On the other hand, the $\tau_s$ and $\tau_b$ masses are given by
\begin{align}
m_{\tau_s}^2 &= \frac{\partial_{\tau_s}^2 V}{2K_{T_s\bar T_s}} \sim \frac{\e^{t/g_s}}{g_s} \left(g_s^2\partial_{t}^2 V-2g_sv\partial_{ t}\partial_{v} V+v^2\partial_{v}^2 V\right) \sim \frac{\e^{t/g_s}}{g_s} \partial_{v}^2 V \sim \e^{-2t/g_s}, \label{gsgtsg} \\
m_{\tau_b}^2 &= \frac{\partial_{\tau_b}^2 V}{2K_{T_b\bar T_b}} \sim \partial_{v}^2 V \sim g_s \e^{-3t/g_s}.
\end{align}
As already noted in Section \ref{setup}, $m_{\tau_s}^2$ agrees with $m_2^2$ up to subleading terms, while $m_{\tau_b}^2$ is parametrically heavier than $m_3^2$. In particular, $m_{\tau_b}^2$ is proportional to $\partial_{v}^2 V$ and therefore, contrary to $m_3^2$, not sensitive to the on-shell cancellation effect in $\partial_{t}^2 V$ and $\partial_{ t}\partial_v V$.

In summary, we have seen that the NPNS is responsible for leading-in-$g_s$ cancellations in the vacuum energy and one of the eigenvalues of the mass matrix.
This is conceptually similar to the ordinary no-scale structure, which implies that $V=\e^\mathcal{K} |D_aW|^2$ at the classical level (where the index $a$ runs over the complex-structure moduli and the axio-dilaton). Since $\e^\mathcal{K}\sim \mathcal{V}^{-2}$ classically, $V$ has a runaway direction unless $D_aW=0$. In the latter case, $V$ is independent of the K\"ahler moduli and yields a family of Minkowski vacua \cite{Giddings:2001yu}. The vacuum energy and the K\"ahler-moduli masses are then corrected due to subleading-in-$\mathcal{V}$ terms breaking the no-scale structure. Analogously, the NPNS implies that \eqref{tv} has a runaway direction at leading order in $g_s$ unless $\frac{4\sqrt{t}}{3v} - \frac{2 t}{v^2} + \frac{3}{8v^3}=0$. In the latter case, $V$ is independent of $t$ and yields a family of Minkowski vacua. The NPNS is again broken by subleading effects, but this time subleading in $g_s$ rather than $\mathcal{V}$.\footnote{Note that the NPNS can be viewed as an example for a ``generalized no-scale structure'' \cite{Barbieri:1985wq, Burgess:2020qsc}, which requires an eigenvalue of the matrix $M_{i\bar\jmath}=\partial_i \partial_{\bar\jmath} \exp \left[-(\mathcal{K}+\ln|W|^2)/3\right]$ to vanish. In the LVS, one of the two eigenvalues of $M_{i\bar\jmath}$ indeed vanishes at leading order in $g_s$ upon substituting the solution.}

An important point is that the NPNS is also broken when we add terms to the potential \eqref{tv} which do not scale like $\e^{-3t/g_s}$.
According to the above discussion, there is no reason why such terms should cancel in $V_0$ and $m_3^2$ at leading order in $g_s$.
We therefore expect them to be \emph{enhanced} in $V_0$ and $m_3^2$ relative to the other terms by a factor $1/g_s$. This is true, for example, for the uplift term in the LVS potential (i.e., the $\zeta$-dependent terms in the second line of \eqref{lvs-potential}), which scales like $\e^{-4t/3g_s}$ and therefore breaks the NPNS. The relative factor between this term and the $\xi$ term in \eqref{lvs-potential} is of the order $g_s\alpha$, as can be checked using \eqref{zeta0} and \eqref{alpha}. On the other hand, as is evident from \eqref{v0} and \eqref{abc}, the relative factor in $V_0$ and $m_3^2$ is $\alpha$ and thus by a factor $1/g_s$ larger.

Crucially, also some of the corrections discussed in Section \ref{vol} break the NPNS. In particular, this is the case for the last term in \eqref{dv} and the terms in \eqref{cflux2}, \eqref{ccon} and \eqref{cf2}.
Indeed, none of these terms scale like $\e^{-3t/g_s}$ in the scalar potential.
If such an NPNS-breaking correction is suppressed by a small factor $\epsilon$ in the off-shell potential, it is only suppressed by a factor $\epsilon/g_s$ in $V_0$ and $m_3^2$.
Corrections which are naively negligible in the off-shell potential for $\epsilon\ll 1$ may therefore still change the sign of the vacuum energy or create an instability unless the stronger condition $\epsilon/g_s \ll 1$ holds.

\section{Effect of Corrections}
\label{problems}

In this section, we discuss how the various corrections of Section \ref{vol} affect the LVS vacua.
In particular, we compute the backreaction of the corrections on the conifold modulus and the K\"ahler moduli. We furthermore point out that the volume and the uplift parameter $\alpha$ are exponentially sensitive to the corrections. We also compute the corrections to the vacuum energy and the moduli masses. We will see that this shifts the boundaries of the $\alpha$ interval \eqref{range} defining the dS region and can even make this region disappear altogether.

\subsection{Conifold Modulus}

We first compute the backreaction of the corrections on the conifold modulus $\zeta$. The relevant corrections are the loop correction in the second line of \eqref{dv} and the $1/(g_sM)^2$ correction in \eqref{ccon}, as all other corrections do not depend on $\zeta$.\footnote{Here and in the following, we assume that $\mathcal{C}_s^\text{KK}$, $\mathcal{C}_b^\text{KK}$ and the various other coefficients do not have a $\zeta$ dependence.}

It is instructive to first discuss the $1/(g_sM)^2$ correction, which has a very simple form. In particular, adding this term to the leading LVS potential \eqref{lvs-potential} effectively replaces $c^{\prime\prime} \to c^{\prime\prime}\left(1+\frac{\mathcal{C}^\text{con}}{(g_sM)^2}\right)$ in \eqref{lvs-potential}. Making the same replacement in \eqref{zeta0}, the $\zeta$ vev becomes
\begin{align}
\zeta &= \e^{ -\frac{2\pi K}{g_sM}-\frac{3}{4}+\sqrt{\frac{9}{16}-\frac{4\pi c^{\prime}c^{\prime\prime}}{g_sM^2}\left(1+\frac{\mathcal{C}^\text{con}}{(g_sM)^2}\right)}}. \label{corr-zeta}
\end{align}
In order for $\zeta$ to have a minimum, the argument of the square root needs to be non-negative. The correction thus modifies the conifold-instability bound  \eqref{gsm}. In particular, the allowed regions for $g_sM^2$ are
\begin{equation}
g_sM^2\ge \frac{32\pi c^\prime c^{\prime\prime}}{9} \left( 1 + \sqrt{1+ \frac{9 \mathcal{C}^\text{con}}{16\pi c^{\prime}c^{\prime\prime} g_s}} \right) \quad \text{or} \quad g_sM^2\le \frac{32\pi c^\prime c^{\prime\prime}}{9} \left( 1 - \sqrt{1+ \frac{9 \mathcal{C}^\text{con}}{16\pi c^{\prime}c^{\prime\prime} g_s}} \right).
\end{equation}
Depending on the value of $\mathcal{C}^\text{con}$, these conditions are stronger or weaker compared to the leading result \eqref{gsm}. For $\mathcal{C}^\text{con}>0$, the first inequality yields a lower bound on $g_sM^2$ which is stronger than \eqref{gsm}. The second inequality cannot be satisfied for $g_sM^2>0$.
On the other hand, for $\mathcal{C}^\text{con}<0$, the lower bound on $g_sM^2$ on the left-hand side is relaxed compared to \eqref{gsm}. Furthermore, the inequality on the right-hand side opens up a new allowed region for $g_sM^2$. For $\mathcal{C}^\text{con}\le - \frac{16\pi c^{\prime}c^{\prime\prime} g_s}{9} $, the two allowed regions merge and the conifold-instability bound becomes trivial.

We now also take into account the loop correction in the second line of \eqref{dv}.
Solving the $\zeta$ equation for the corrected potential, we find
\begin{align}
\zeta &= \left(1- \mathcal{C}^\text{con} \frac{8\pi c^\prime c^{\prime\prime}}{g_s^{3}M^4\sqrt{9-\frac{64\pi c^\prime c^{\prime\prime}}{g_sM^2}}}\right) \e^{-\frac{2\pi K}{g_sM}-\frac{3}{4}+\sqrt{\frac{9}{16}-\frac{4\pi c^{\prime}c^{\prime\prime}}{g_sM^2}}} \nl
-\mathcal{C}^\text{KK}_b\frac{24\pi^2 c^{\prime 2}g_s^3 M^2 |W_0|^2 }{\sqrt{9-\frac{64\pi c^\prime c^{\prime\prime}}{g_sM^2}}\,\mathcal{V}^2  } \, \e^{-\frac{2\pi K}{3g_sM}-\frac{1}{4}+\sqrt{\frac{1}{16}-\frac{4\pi c^{\prime}c^{\prime\prime}}{9g_sM^2}}}. \label{corr-zeta2}
\end{align}
Here and in the following, we restrict to expressions at the linear order in the corrections.\footnote{There is no benefit in keeping higher than linear terms, as these would compete with higher-order corrections to the scalar potential, which we have not computed in Section \ref{vol}.}
Note that both corrections have a $\sqrt{9-\frac{64\pi c^\prime c^{\prime\prime}}{g_sM^2}}$ factor in their denominators and thus blow up as we approach the naive conifold-instability bound \eqref{gsm}. This reflects the fact that the corrections change the dynamics of the conifold modulus in the vicinity of this bound, as discussed above.

\subsection{K\"ahler Moduli and Uplift Parameter}
\label{mi}

We now turn to the K\"ahler moduli. At linear order in the corrections, the solution for $\mathcal{V}$ and $\tau_s$ is
\begin{align}
\mathcal{V} &= \left(\frac{3(a_s\tau_s-1)}{4a_s\tau_s-1} + \mathcal{C}_s^\text{KK}\frac{g_s}{8 \kappa_s \tau_s} - \mathcal{C}_s^\text{log}\frac{3 a_s}{8}\right) \frac{\kappa_s\sqrt{\tau_s}|W_0|}{a_s|A_s|}\,\e^{a_s\tau_s-a_s\chi_s/24g_s}, \label{vol-corr} \\
\tau_s &= \frac{\hat\xi^{2/3}}{(2\kappa_s)^{2/3}g_s} + \frac{1}{3a_s}+\frac{4\alpha}{15a_s}
+ \frac{(425+80\alpha+32\alpha^2)(2\kappa_s)^{2/3}}{1800a_s^2 \hat\xi^{2/3}}g_s
+\mathcal{O}(g_s^2) \nl
- \mathcal{C}_s^\text{KK}\frac{g_s}{3\kappa_s}
+ \mathcal{C}_b^\text{KK}  \frac{32c^\prime}{27(2\kappa_s)^{2/3}\hat\xi^{1/3}}
\frac{g_s^{5/2} M^2 \zeta^{2/3}}{\mathcal{V}^{1/3} } \left(4 + \frac{3}{\sqrt{9-\frac{64\pi c^\prime c^{\prime\prime}}{g_sM^2}}} \right)
+ \mathcal{C}_s^\text{log}\frac{\hat\xi^{2/3} a_s }{(2\kappa_s)^{2/3}g_s} \nl + \mathcal{C}_1^\xi\frac{2 \hat\xi^{1/3} a_s }{3(2\kappa_s)^{4/3}g_s}
+\mathcal{C}_2^\xi\frac{2}{3(2\kappa_s)^{2/3}\hat\xi^{1/3}}
+ \mathcal{C}^\text{flux}\frac{110 }{27 (2\kappa_s)^{2/3}\hat\xi^{1/3}} \frac{KM}{g_s\mathcal{V}^{2/3}} \nl
+\mathcal{C}^\text{con} \frac{2\alpha}{15a_s g_s^2M^2} \left(1+ \frac{3}{\sqrt{9-\frac{64\pi c^\prime c^{\prime\prime}}{g_sM^2}}} \right)
+ \mathcal{C}^F\frac{176}{81 (2\kappa_s)^{2/3}\hat\xi^{1/3}} \frac{|W_0|^2}{\mathcal{V}^{2/3}}. \label{taus-corr}
\end{align}
Here $\hat\xi=\xi-\Delta\xi$ and we only displayed the leading order in $g_s$ in each of the correction terms $\sim \mathcal{C}_\circ^{\circ}$.\footnote{For the computation of $V_0$ and $m_3^2$ in the next subsection, one also needs to take into account the next-to-leading terms in $g_s$ since some of the leading terms cancel out due to the NPNS.} The $\mathcal{C}^\text{log}_s$ terms in $\mathcal{V}$ and $\tau_s$ are consistent with a result of \cite{Conlon:2010ji} (obtained there for an LVS potential without uplift term).

The approximation of
neglecting the
$\mathcal{C}_\circ^{\circ}$ terms in \eqref{vol-corr}, \eqref{taus-corr} breaks down if they are $\mathcal{O}(1)$ or larger. Because of $\mathcal{V} \sim \e^{a_s\tau_s}$, the volume then differs by an exponentially large factor from the uncorrected value computed in Section \ref{setup}.
The same is true for the uplift parameter $\alpha \sim \mathcal{V}^{5/3}$, which determines whether an extremum corresponds to a dS vacuum (see Section \ref{setup}).
Therefore, unless all of the above corrections are small, the properties of an extremum of the corrected potential are very different from those of the corresponding extremum of the uncorrected potential. In particular, naive dS vacua can in reality be AdS vacua or unstable (and vice versa).

This problem is exacerbated by the fact that a correction which is small in the off-shell LVS potential need not be small in \eqref{taus-corr}. The reason is that the leading-order solution for $\tau_s$ scales likes $1/g_s$.
An $\mathcal{O}(\epsilon)$ correction to the $\tau_s$ potential therefore corrects \eqref{taus-corr} by a term of the order $\epsilon/g_s$.
Due to this $1/g_s$ enhancement, some of the $\mathcal{C}_\circ^\circ$ terms in \eqref{taus-corr} have no parametric suppression at all or are even parametrically large.

In particular, the $\mathcal{C}_s^\text{log}$ and $\mathcal{C}_1^\xi$ terms in \eqref{taus-corr} scale like $1/g_s$ and thus blow up at small coupling.
Also the $\mathcal{C}^\xi_2$ term is dangerous, as it is not suppressed by $g_s$ or any other small parameter.
It is therefore in general not self-consistent to neglect these terms.
One might wonder whether
the parameters $\kappa_s$, $a_s$ and $\hat\xi$ or the coefficients $\mathcal{C}_s^\text{log}$ and $\mathcal{C}_i^\xi$ could produce small factors that help to suppress the dangerous terms. While we cannot exclude that this is true in special models, it is not clear why such a property should be expected in general. Indeed, $\kappa_s$, $a_s$ and $\hat\xi$ are $\mathcal{O}(1)$ numbers fixed by the geometry and brane data (see, e.g., Section \ref{dsmin} for an explicit CY model), and the coefficients $\mathcal{C}_s^\text{log}$ and $\mathcal{C}_i^\xi$ are also expected to be $\mathcal{O}(1)$ generically.

Let us also discuss the remaining corrections in \eqref{taus-corr}, i.e., the terms proportional to $\mathcal{C}_s^\text{KK}$, $\mathcal{C}_b^\text{KK}$, $\mathcal{C}^\text{flux}$, $\mathcal{C}^\text{con}$ and $\mathcal{C}^F$. These depend on the tunable parameters $g_s$, $W_0$, $K$ and $M$ as well as on $\zeta$ and $\mathcal{V}$ (which are functions of these parameters). Each correction can thus in principle be made small by an appropriate parameter choice. For example, the $\mathcal{C}_s^\text{KK}$ correction becomes negligible for small enough $g_s$, the $\mathcal{C}^\text{flux}$ correction for large enough $\mathcal{V}$, etc.
One might therefore hope to identify models in which the $\mathcal{C}_s^\text{log}$ and $\mathcal{C}_i^\xi$ corrections are harmless for some reason (e.g., because the coefficients happen to vanish) and then find a point in the $(g_s, W_0, K,M)$ parameter space where all other corrections are suppressed. However, we will show in Section \ref{dsmin} in an explicit model that this does not work. Indeed, whenever the vacuum energy is positive, it is impossible to make all of these further corrections small \emph{at the same time}. Hence, even if the naively most dangerous terms $\mathcal{C}_s^\text{log}$, $\mathcal{C}_1^\xi$ and  $\mathcal{C}_2^\xi$ could be controlled somehow, one would still face the problem of a number of other large corrections.

As explained above, this implies an exponentially large uncertainty in the vev of the volume modulus and the uplift parameter at any point in the parameter space for which the uncorrected potential yields dS vacua. In other words, it is impossible to determine where dS vacua are located in the parameter space without computing the coefficients of the above corrections explicitly.

\subsection{Vacuum Energy and Moduli Masses}
\label{md}

Aside from backreacting on the moduli vevs, the corrections of Section \ref{vol}
also shift the boundaries of the $\alpha$ interval \eqref{range} that defines the dS region.
Recall that these boundaries are determined by demanding that the on-shell potential $V_0$ and the eigenvalue $m_3^2$ of the Hessian are positive.
Repeating the computation of Section \ref{setup} including the various corrections of Section \ref{vol}, we find\footnote{We again consider a field basis which is canonically normalized up to terms subleading in powers of $1/\mathcal{V}$. This is achieved with the same transformation matrix \eqref{m0} as in Section \ref{setup}, except that the upper left entry is replaced by $\frac{2\tau_s^{1/4}\sqrt{\mathcal{V}}}{\sqrt{3}\sqrt{\kappa_s}} +\frac{ \mathcal{C}^\text{KK}_s g_s\sqrt{\mathcal{V}}}{6\sqrt{3}\tau_s^{3/4}\kappa_s^{3/2}} - \frac{\mathcal{C}^\text{log}\ln\mathcal{V}\sqrt{\mathcal{V}}}{2\sqrt{3}\tau_s^{3/4}\sqrt{\kappa_s}}$.}
\begin{equation}
V_0 = \frac{3(2\kappa_s)^{2/3}\hat\xi^{1/3} \sqrt{g_s}|W_0|^2}{16a_s \mathcal{V}^{3}} \rho, \qquad m_3^2 = \frac{9(2\kappa_s)^{2/3}\hat\xi^{1/3} \sqrt{g_s}|W_0|^2}{8a_s \mathcal{V}^{3}} \mu_3 \label{rhomu3}
\end{equation}
with
\begin{align}
\rho &= \alpha-1 + \frac{(10+\alpha+4\alpha^2)(2\kappa_s)^{2/3}}{30a_s\hat\xi^{2/3}}g_s + \mathcal{O}(g_s^2) \nl
-\mathcal{C}^\text{KK}_s \frac{\alpha}{3(2\kappa_s)^{1/3}\hat\xi^{2/3}}g_s^2
-\mathcal{C}^\text{KK}_b \frac{8 c^\prime a_s}{9(2\kappa_s)^{2/3}\hat\xi^{1/3}}\frac{g_s^{5/2} M^2 \zeta^{2/3}}{\mathcal{V}^{1/3}}
\left( 7-\frac{15}{\sqrt{9-\frac{64\pi c^\prime c^{\prime\prime}}{g_sM^2}}} \right) \nl
+\mathcal{C}^\text{log}_s \frac{(2+\alpha) a_s}{2}
+\mathcal{C}^\xi_1 \frac{(1+\alpha)a_s}{3(2\kappa_s)^{2/3}\hat\xi^{1/3}}
- \mathcal{C}^\xi_2 \frac{(1-\alpha)}{3\hat\xi } g_s
- \mathcal{C}^\text{flux} \frac{20 a_s}{9(2\kappa_s)^{2/3}\hat\xi^{1/3}} \frac{KM}{g_s\mathcal{V}^{2/3}} \nl
+\mathcal{C}^\text{con} \frac{\alpha}{2g_s^2M^2}\left( 1+\frac{3}{\sqrt{9-\frac{64\pi c^\prime c^{\prime\prime}}{g_sM^2}}} \right)
- \mathcal{C}^F \frac{32 a_s}{27(2\kappa_s)^{2/3}\hat\xi^{1/3}} \frac{|W_0|^2}{\mathcal{V}^{2/3}} , \label{alpharange1} \\
\mu_3 &= \frac{9}{4} -\alpha
- \frac{(135-16\alpha+16\alpha^2)(2\kappa_s)^{2/3}}{120a_s\hat\xi^{2/3} }g_s + \mathcal{O}(g_s^2) \nl
+\mathcal{C}^\text{KK}_s \frac{\alpha}{3(2\kappa_s)^{1/3}\hat\xi^{2/3}}g_s^2
+\mathcal{C}^\text{KK}_b \frac{40 c^\prime a_s}{9(2\kappa_s)^{2/3}\hat\xi^{1/3}}\frac{g_s^{5/2} M^2 \zeta^{2/3}}{\mathcal{V}^{1/3}}
\left( 2-\frac{3}{\sqrt{9-\frac{64\pi c^\prime c^{\prime\prime}}{g_sM^2}}} \right) \nl
-\mathcal{C}^\text{log}_s \frac{(9+2\alpha) a_s}{4}
-\mathcal{C}^\xi_1 \frac{(9+4\alpha)a_s}{12(2\kappa_s)^{2/3}\hat\xi^{1/3}}
+ \mathcal{C}^\xi_2 \frac{9-4\alpha}{12 \hat\xi} g_s
+\mathcal{C}^\text{flux} \frac{55 a_s}{9(2\kappa_s)^{2/3}\hat\xi^{1/3}} \frac{KM}{g_s\mathcal{V}^{2/3}} \nl
-\mathcal{C}^\text{con} \frac{\alpha}{2g_s^2M^2}\left( 1+\frac{3}{\sqrt{9-\frac{64\pi c^\prime c^{\prime\prime}}{g_sM^2}}} \right)
+ \mathcal{C}^F \frac{88 a_s}{27(2\kappa_s)^{2/3}\hat\xi^{1/3}} \frac{|W_0|^2}{\mathcal{V}^{2/3}} , \label{alpharange2}
\end{align}
where $\hat\xi=\xi-\Delta\xi$ as before. The first lines in \eqref{alpharange1} and \eqref{alpharange2} are obtained when the LVS potential of Section \ref{setup} is assumed (up to the replacement $\xi\to\hat\xi$). The remaining terms in both equations are due to the corrections of Section \ref{vol}, where we restricted to the linear order in $\mathcal{C}^\circ_\circ$ and only displayed the leading-in-$g_s$ term for each correction.

Solving the equations $\rho(\alpha)=0$ and $\mu_3(\alpha)=0$ yields the lower and upper bounds $\alpha=\alpha_\text{low}$ and $\alpha=\alpha_\text{up}$, respectively, which determine the $\alpha$ interval corresponding to dS vacua. Ignoring the $\mathcal{C}^\circ_\circ$ terms, one finds $\alpha_\text{low}=1$, $\alpha_\text{up}=\frac{9}{4}$ (at leading order in $g_s$) and thus a non-empty dS region. However, this region is clearly in danger of vanishing if some of the $\mathcal{C}^\circ_\circ$ terms in \eqref{alpharange1} and \eqref{alpharange2} are $\mathcal{O}(1)$.

In particular, the $\mathcal{C}_s^\text{log}$ and $\mathcal{C}^\xi_1$ terms are not suppressed by any small parameters in \eqref{alpharange1} and \eqref{alpharange2} and thus among the most dangerous terms. In addition, they affect \eqref{alpharange1} and \eqref{alpharange2} indirectly by backreacting on $\mathcal{V}$ and $\alpha$, as explained in Section \ref{mi}. Each of the remaining corrections in \eqref{alpharange1} and \eqref{alpharange2} can be made small individually by an appropriate choice of the parameters $g_s$, $W_0$, $K$ and $M$. However, as noted before, we will show in Section \ref{dsmin} in an explicit model that there is no point in the parameter space where the vacuum energy is positive and all corrections are small at the same time.
Hence, even if one assumes $\mathcal{C}_s^\text{log}=\mathcal{C}^\xi_1=0$, the corrections to $V_0$ and $m_3^2$ are $\mathcal{O}(1)$.

An important related observation is that the $\mathcal{C}^\text{KK}_b$, $\mathcal{C}^\text{flux}$, $\mathcal{C}^\text{con}$ and $\mathcal{C}^F$ terms in \eqref{alpharange1} and \eqref{alpharange2} are parametrically larger than one might have expected from their suppression in the off-shell potential.
For example, comparing \eqref{cflux2} with \eqref{lvs-potential}, one finds that the $\mathcal{C}^\text{flux}$ term is suppressed by a factor $KM/\mathcal{V}^{2/3}$ in the off-shell potential. One might thus falsely conclude that the correction can be ignored whenever $KM/\mathcal{V}^{2/3}\ll 1$. However, in \eqref{alpharange1} and \eqref{alpharange2}, the $\mathcal{C}^\text{flux}$ term is only suppressed by a factor $KM/g_s\mathcal{V}^{2/3}$, which is parametrically larger by a factor $1/g_s$. One can verify that the same $1/g_s$ enhancement occurs for the $\mathcal{C}^\text{KK}_b$, $\mathcal{C}^\text{con}$ and $\mathcal{C}^F$ terms as well (see Table \ref{tab-corr}).
As explained in Section \ref{npns}, this is due to the fact that these corrections break the NPNS.

For completeness, we also state the corrected expressions for the other two mass eigenvalues $m_1^2$ and $m_2^2$. When they are computed using the uncorrected LVS potential \eqref{lvs-potential}, they are manifestly positive (for small $g_s$ and large $\mathcal{V}$) and thus do not yield any non-trivial constraints. However, including the corrections of Section \ref{vol}, this is no longer true. In particular, we find
\begin{equation}
m_1^2 = \frac{3}{32\pi^2 c^\prime} \sqrt{9-\frac{64\pi c^\prime c^{\prime\prime}}{g_sM^2}} \frac{4\zeta^{2/3}}{9c^\prime g_sM^2 \mathcal{V}^{2/3}} \mu_1, \qquad m_2^2 = \frac{2 a_s^2\hat\xi^{4/3}|W_0|^2}{(2\kappa_s)^{4/3}g_s\mathcal{V}^2}\mu_2
\end{equation}
with
\begin{align}
\mu_1 &= 1+\mathcal{C}^\text{KK}_b \frac{20 c^\prime a_s}{9\alpha (2\kappa_s)^{2/3}\hat \xi^{1/3}}\frac{g_s^{5/2} M^2 \zeta^{2/3}}{\mathcal{V}^{1/3}}
\left( 1 +\frac{3}{\sqrt{9-\frac{64\pi c^\prime c^{\prime\prime}}{g_sM^2}}} - \frac{18}{9-\frac{64\pi c^\prime c^{\prime\prime}}{g_sM^2}} \right) \nl
+\mathcal{C}^\text{con} \frac{1}{2g_s^2 M^2}\left( 1-\frac{9}{9-\frac{64\pi c^\prime c^{\prime\prime}}{g_sM^2}} \right)
, \label{mu1} \\
\mu_2 &= 1 - \frac{(35-32\alpha)(2\kappa_s)^{2/3}}{60a_s\hat\xi^{2/3}}g_s + \mathcal{O}(g_s^2) \nl
-\mathcal{C}^\text{KK}_s \frac{8\alpha (2\kappa_s)^{1/3}}{45a_s\hat\xi^{4/3}}g_s^3
+\mathcal{C}^\text{KK}_b \frac{64 c^\prime}{27\hat \xi}\frac{g_s^{7/2} M^2 \zeta^{2/3}}{\mathcal{V}^{1/3}}
\left( 4+\frac{3}{\sqrt{9-\frac{64\pi c^\prime c^{\prime\prime}}{g_sM^2}}} \right) \nl
+\mathcal{C}^\text{log}_s \frac{2(2\kappa_s)^{2/3}(2\alpha-35)}{15\hat\xi^{2/3}}g_s 
+\mathcal{C}^\xi_1 \frac{4 a_s}{3(2\kappa_s)^{2/3}\hat\xi^{1/3}}
+ \mathcal{C}^\xi_2 \frac{4}{3\hat\xi } g_s
+ \mathcal{C}^\text{flux} \frac{220}{27 \hat\xi} \frac{KM}{\mathcal{V}^{2/3}} \nl
+\mathcal{C}^\text{con} \frac{4\alpha (2\kappa_s)^{2/3}}{15 a_s\hat\xi^{2/3}} \frac{1}{g_s M^2}\left( 1+\frac{3}{\sqrt{9-\frac{64\pi c^\prime c^{\prime\prime}}{g_sM^2}}} \right)
+ \mathcal{C}^F \frac{352}{81\hat\xi} \frac{g_s|W_0|^2}{\mathcal{V}^{2/3}}. \label{mu2}
\end{align}
Here we again restricted to the linear order in the $\mathcal{C}^\circ_\circ$ corrections and only displayed the leading-in-$g_s$ term for each correction. We thus see that the corrections that affect $V_0$ and $m_3^2$ are potentially also dangerous for $m_1^2$ and $m_2^2$.\footnote{Note that the $\mathcal{C}^\text{KK}_b$, $\mathcal{C}^\text{flux}$, $\mathcal{C}^\text{con}$ and $\mathcal{C}^F$ terms in \eqref{mu2} are suppressed by an extra factor $g_s$ compared to those in \eqref{alpharange1} and \eqref{alpharange2}. As explained in Section \ref{npns}, this is related to the fact that $m_2^2$ is not sensitive to the NPNS cancellations.
One furthermore observes that the $\mathcal{C}^\text{KK}_s$ and $\mathcal{C}^\text{log}_s$ corrections only appear at the relative orders $g_s^3$ and $g_s$ in $m_2^2$, whereas they appear at the relative orders $g_s^2$ and $g_s^0$ in \eqref{alpharange1} and \eqref{alpharange2}. The corrections in $m_2^2$ are thus somewhat milder than in $V_0$ and $m_3^2$.
}

\begin{table}[t]\renewcommand{\arraystretch}{1.2}\setlength{\tabcolsep}{6pt}
\hspace{-0.85cm}
\scriptsize
\begin{tabular}{|p{3.4cm}|p{1.2cm} p{0.9cm} p{2.4cm} p{3.3cm} p{3.3cm}| }
\hline 
type of correction & coefficient & breaks NPNS & \mbox{suppression in}  \mbox{off-shell potential} &  \mbox{suppression in exponent} \mbox{of $\mathcal{V}$} &  \mbox{suppression in $\rho$} \mbox{and $\mu_3$} \\
\hline 
$g_s$ corrections ($\alpha^{\prime 2}$) & $\mathcal{C}_s^\text{KK}$ & no & $g_s^2$ & $g_s$ & $g_s^2$ \\
 & $\mathcal{C}_b^\text{KK}$ & yes & $g_s^{7/2}M^2\zeta^{2/3}/\mathcal{V}^{1/3}$ & $g_s^{5/2}M^2\zeta^{2/3}/\mathcal{V}^{1/3}$ & $g_s^{5/2}M^2\zeta^{2/3}/\mathcal{V}^{1/3} $ \\
log field redefinitions & $\mathcal{C}^\text{log}_s$ & no & none & $1/g_s$  & none \\
O7 self-intersection & $\Delta\xi$ & no & none & appears in various terms & appears in various terms \\
$g_s$ corrections ($\alpha^{\prime 3}$) & $\mathcal{C}^\xi_1$  & no & none & $1/g_s$ & none \\
 & $\mathcal{C}^\xi_{2}$  & no & $g_s$ & none & $g_s$ \\
gauge-kinetic function & $\chi_s$  & no & none & $1/g_s$ & no explicit dependence \\
conifold-flux backreaction & $\mathcal{C}^\text{flux}$  & yes & $KM/\mathcal{V}^{2/3}$ & $KM/g_s\mathcal{V}^{2/3}$ & $KM/g_s\mathcal{V}^{2/3}$ \\
conifold curvature & $\mathcal{C}^\text{con}$  & yes & $1/g_sM^2$ & $1/(g_sM)^2 $  & $1/(g_sM)^2 $ \\
higher $F$ terms & $\mathcal{C}^F$  & yes & $g_s|W_0|^2/\mathcal{V}^{2/3}$ & $|W_0|^2/\mathcal{V}^{2/3}$ & $|W_0|^2/\mathcal{V}^{2/3}$ \\
\hline 
\end{tabular}
\caption{Parametric suppression of different corrections to the LVS potential in a dS vacuum (ignoring $\mathcal{O}(1)$ factors and the dependence on $\sqrt{9-\frac{64\pi c^\prime c^{\prime\prime}}{g_sM^2}}$).
The suppression in the off-shell potential is stated relative to the $\xi$ term and near the minimum.\\
}
\label{tab-corr}
\end{table}

\subsection{Conditions for Perturbative Control}
\label{pert}

Let us summarize the results of the preceding sections. We have seen that the LVS potential receives a variety of corrections that can affect its minima.
In particular, as stated in \eqref{vol-corr}--\eqref{mu2}, the corrections appear in the exponent of the vev of the volume modulus, in the on-shell potential and in the eigenvalues of the mass matrix.

The least dangerous corrections are the $\Delta\xi$ and $\chi_s$ corrections of Sections \ref{corr1} and \ref{corr2}, as their appearance in the relevant expressions is fully understood including numerical factors. We stress, however, that it is in general not self-consistent to ignore these terms. For example, they appear in the volume through an exponential factor $\mathcal{V}\sim\e^{a_s\left[ \hat\xi^{2/3}/(2\kappa_s)^{2/3}g_s - \chi_s/24g_s\right]}$ and thus typically have a large effect. However, since we know how these terms enter the potential, we can easily take them into account in any computation. On the other hand, all other corrections of Section \ref{vol} are only partially known in the sense that we expect them to exist generically but do not know their precise numerical coefficients $\mathcal{C}^\circ_\circ$. We therefore need to ensure that these corrections can be neglected self-consistently.

Unfortunately, the $\mathcal{C}^\text{log}_s$ and $\mathcal{C}^\xi_i$ corrections are not suppressed by small parameters such as $g_s$ or $1/\mathcal{V}$ in the relevant expressions.
In particular, for perturbative control of $\mathcal{V}\sim \e^{a_s\tau_s}$, we require from \eqref{taus-corr} that
\begin{equation}
\frac{\hat\xi^{2/3} a_s^2 |\mathcal{C}_s^\text{log}|}{(2\kappa_s)^{2/3}},\frac{2 \hat\xi^{1/3} a_s^2 |\mathcal{C}_1^\xi|}{3(2\kappa_s)^{4/3}}\ll g_s, \qquad
\frac{2a_s|\mathcal{C}_2^\xi|}{3(2\kappa_s)^{2/3}\hat\xi^{1/3}} \ll 1. \label{c-cond}
\end{equation}
It is not obvious that these conditions can be satisfied in explicit models since all numbers on the left-hand sides are expected to be $\mathcal{O}(1)$.
Instead of searching for models that satisfy \eqref{c-cond}, one could (in principle) also try to take into account the full non-linear $\mathcal{C}^\text{log}_s$ and $\mathcal{C}^\xi_i$ corrections in the relevant expressions and look for dS vacua in such a modified scenario. In either case, the coefficients $\mathcal{C}^\text{log}_s$ and $\mathcal{C}^\xi_i$ have to be computed explicitly, which is a formidable task.
We will not have anything further to say about this problem in the remainder of this paper. Instead, we will focus on the other types of corrections from now on, which are problematic in their own right.

Indeed, even assuming that \eqref{c-cond} is satisfied in a given model, we still need to control the remaining corrections.
Their suppression factors in the relevant expressions can be read off of \eqref{corr-zeta2}--\eqref{mu2}. In particular, $V_0$ and $m_3^2$ yield strong constraints. Assuming $\mathcal{C}^\circ_\circ=\mathcal{O}(1)$ and requiring small corrections in \eqref{alpharange1}, \eqref{alpharange2} compared to the leading terms $\alpha-1$ and $\frac{9}{4}-\alpha$, we can write the
necessary conditions for perturbative control as\footnote{Here we assume that the dominant terms in \eqref{alpharange1}, \eqref{alpharange2} are $\alpha-1$ and $\frac{9}{4}-\alpha$, respectively. There is also a second regime, where $\alpha$ is very close to either $1$ or $\frac{9}{4}$ such that the $\mathcal{O}(g_s)$ terms in the first lines of \eqref{alpharange1}, \eqref{alpharange2} are dominant. However, in that case, it would be parametrically harder to suppress the $\mathcal{C}^\circ_\circ$ corrections so that we do not consider this case separately.}
\begin{equation}
\lambda_i \ll 1 \label{lambda}
\end{equation}
with
\begin{align}
 \lambda_1&\equiv \text{max}\left(\frac{10+\alpha+4\alpha^2}{30|\alpha-1|}, \frac{135-16\alpha+16\alpha^2}{120|\frac{9}{4}-\alpha|} \right) \frac{(2\kappa_s)^{2/3}}{a_s\hat\xi^{2/3}} g_s, \label{lambda00} \\
 \lambda_2 &\equiv \text{max}\left[\frac{1}{|\alpha-1|} \left|7-\frac{15}{\sqrt{9-\frac{64\pi c^\prime c^{\prime\prime}}{g_sM^2}}}\right|,\frac{1}{|\frac{9}{4}-\alpha|} \left|10-\frac{15}{\sqrt{9-\frac{64\pi c^\prime c^{\prime\prime}}{g_sM^2}}}\right| \right] \nl \times \frac{8c^\prime a_s}{9(2\kappa_s)^{2/3}\hat\xi^{1/3}} \frac{g_s^{5/2}M^2\zeta^{2/3}}{\mathcal{V}^{1/3}}, \\
 \lambda_3 &\equiv \text{max}\left(\frac{20}{|\alpha-1|},\frac{55}{|\frac{9}{4}-\alpha|} \right) \frac{a_s}{9(2\kappa_s)^{2/3}\hat\xi^{1/3}} \frac{KM}{g_s\mathcal{V}^{2/3}}, \\
 \lambda_4 &\equiv \text{max}\left(\frac{1}{|\alpha-1|},\frac{1}{|\frac{9}{4}-\alpha|} \right) \frac{\alpha}{2(g_sM)^2} \left(1+ \frac{3}{\sqrt{9-\frac{64\pi c^\prime c^{\prime\prime}}{g_sM^2}}}\right), \\
 \lambda_5 &\equiv \text{max}\left(\frac{32}{|\alpha-1|},\frac{88}{|\frac{9}{4}-\alpha|} \right)\frac{a_s}{27(2\kappa_s)^{2/3}\hat\xi^{1/3}} \frac{|W_0|^2}{\mathcal{V}^{2/3}}. \label{lambda0}
\end{align}
Here $\text{max}(a,b)=a$ for $a\ge b$ and $\text{max}(a,b)=b$ otherwise.

We emphasize that a violation of \eqref{lambda}
does not prove that a candidate dS minimum does not exist. Rather, the conditions provide an estimate for the degree of control in situations where the full solution including all warping and string corrections is not known explicitly.
In principle, one could also imagine a model in which the corrections happen to be subleading in spite of naively too-large $\lambda_i$. For example, some of the numerical coefficients $\mathcal{C}^\circ_\circ$, which are generically expected to be $\mathcal{O}(1)$, could be small or even vanish in certain models. However, the point is again that this would have to be verified by computing these coefficients explicitly. For example, in order to fix $\mathcal{C}^\text{flux}$, one would have to compute the warp factor on the CY (i.e., the Green's function of the Laplacian) and perform a dimensional reduction of various $\alpha^\prime$ corrections to the 10D supergravity action.
On the other hand, in the regime \eqref{lambda}, all potentially dangerous corrections can be self-consistently neglected even when we are \emph{not} able to compute them. We will therefore adopt \eqref{lambda} as a necessary condition for control. Conversely, we will regard vacua which are not self-consistent in this sense as being in the swampland.

As we will see below, \eqref{lambda} can in fact not be satisfied in the LVS. In particular, we will show that, in \emph{every} dS vacuum of the explicit CY model of \cite{Crino:2020qwk}, at least one of the $\lambda_i$ is $\gtrsim\mathcal{O}(1)$.

\section{An Explicit Model}

\label{dsmin}

In this section, we study our claims in an explicit CY compactification, which was previously analyzed in \cite{Crino:2020qwk} and is based on a manifold in the database of \cite{Altman:2014bfa}. We refer to \cite{Crino:2020qwk} for all details about the model and only state what is necessary for our arguments here. The relevant parameters are
\begin{gather}
a_s=\frac{\pi}{3}, \qquad\! \kappa_s=\frac{\sqrt{2}}{9}, \qquad\! \xi= \frac{130\zeta(3)}{(2\pi)^3}, \qquad\! \Delta\xi = \frac{18\zeta(3)}{(2\pi)^3}, \qquad\! \chi_s=3, \qquad\! Q_3=149, \label{parameters}
\end{gather}
where $Q_3$ denotes the D3 tadpole. Following \cite{Crino:2020qwk}, we will furthermore set $A_s=1$. The free parameters in this model are thus $g_s$, $W_0$, $K$ and $M$, where $K$ and $M$ are positive integers with $KM\le Q_3$.\footnote{Note that $g_s$ and $W_0$ (and $A_s$)
are fixed by the stabilization of the complex-structure moduli and the dilaton in an actual string compactification.
Treating $g_s$ and $W_0$ as free continuous parameters should be justified under the usual assumption of a flux landscape admitting a huge number of different solutions.
We will assume that this is the case (see, however, \cite{Braun:2020jrx, Bena:2020xrh}).}

\subsection{Analysis of the Minimum Found in \cite{Crino:2020qwk}}
\label{expl}

Let us illustrate some of the points made in the previous sections in a concrete solution. In particular, \cite{Crino:2020qwk} found a dS minimum for
\begin{equation}
g_s = 0.228, \qquad W_0=23, \qquad K=4, \qquad M=22. \label{parameters2}
\end{equation}
Minimizing the LVS potential \eqref{lvs-potential} using \eqref{parameters} and \eqref{parameters2}, one finds that the moduli vevs and the uplift parameter are
\begin{equation}
\mathcal{V} = 1.87\cdot 10^4,\qquad \tau_s = 7.61, \qquad \zeta= 5.58\cdot 10^{-3}, \qquad \alpha=1.11. \label{ex}
\end{equation}
Since $1<\alpha<\frac{9}{4}$, this indeed corresponds to a dS minimum of the uncorrected potential \eqref{lvs-potential}.

We now check how this result is affected by the various corrections of Section \ref{vol}. We first only turn on the $\Delta\xi$ and $\chi_s$ corrections. 
These are easily taken into account by replacing $\xi\to\xi-\Delta\xi$ and $A_s\to A_s\e^{a_s\chi_s/24g_s}$ in \eqref{lvs-potential}. Minimizing the corrected potential, we find
\begin{equation}
\mathcal{V} = 3.69\cdot 10^3, \qquad \tau_s = 6.69, \qquad \zeta= 5.58\cdot 10^{-3}, \qquad \alpha = 7.89\cdot 10^{-2}. \label{tau}
\end{equation}
The volume is therefore roughly $\frac{1}{5}$ of the value obtained in \eqref{ex}. This reflects the fact that the $\Delta\xi$ and $\chi_s$ terms appear in the exponent in the vev of the volume modulus and can thus have a rather large effect. We also observe that $\alpha$ is by a factor 14 smaller than in \eqref{ex} and thus outside of the range \eqref{range} required for a dS minimum. The corrected minimum is therefore AdS.

Nevertheless, as noted before, the $\Delta\xi$ and $\chi_s$ terms are not really dangerous for the LVS, as it is straightforward to perform a new dS search including these two corrections (as we will indeed do further below). However, this is not possible for the other corrections of Section \ref{vol}, as their coefficients $\mathcal{C}^\circ_\circ$ are not known explicitly. We therefore need to make sure that these corrections are negligible. Using \eqref{lambda00}--\eqref{lambda0}, we find that the parameters controlling the corrections are
\begin{equation}
\lambda_1 = 7.77\cdot 10^{-2}, \qquad \lambda_2 = 0.113, \qquad \lambda_3=12.6, \qquad \lambda_4 =3.93\cdot 10^{-3}, \qquad \lambda_5=9.24 \label{marg}
\end{equation}
at the solution. This violates the condition \eqref{lambda} and thus indicates large unknown corrections to the moduli vevs, the vacuum energy and the moduli masses. In particular, the large values of $\lambda_3$ and $\lambda_5$ imply large $\mathcal{C}^\text{flux}$ and $\mathcal{C}^F$ corrections. As explained in Section \ref{problems}, we furthermore expect large $\mathcal{C}_s^\text{log}$ and $\mathcal{C}^\xi_i$ corrections. Indeed, \eqref{taus-corr}--\eqref{alpharange2} yields
\begin{align}
\tau_s &= 6.69 -0.484 \mathcal{C}^\text{KK}_s+0.482 \mathcal{C}^\text{KK}_b+ 6.61 \mathcal{C}^\text{log}_s +11.7 \mathcal{C}^\xi_1 + 1.77\mathcal{C}^\xi_2+17.5\mathcal{C}^\text{flux} + 9.23 \cdot 10^{-4} \mathcal{C}^\text{con} \nl + 12.8\mathcal{C}^F, \\
V_0 &= -2.96\cdot 10^{-10} \left( 1+ 3.47\cdot 10^{-3}\mathcal{C}^\text{KK}_s+ 3.65\cdot 10^{-2} \mathcal{C}^\text{KK}_b- 1.25 \mathcal{C}^\text{log}_s -1.15 \mathcal{C}^\xi_1 +0.148 \mathcal{C}^\xi_2 \right. \nl \left. + 11.5\mathcal{C}^\text{flux}- 4.16\cdot 10^{-3}\mathcal{C}^\text{con} + 8.39\mathcal{C}^F\right), \\
m_3^2 &= 4.09\cdot 10^{-9}\left( 1+ 1.51\cdot 10^{-3}\mathcal{C}^\text{KK}_s+ 0.123 \mathcal{C}^\text{KK}_b -1.20 \mathcal{C}^\text{log}_s -1.08 \mathcal{C}^\xi_1 + 0.152\mathcal{C}^\xi_2+13.7 \mathcal{C}^\text{flux}  \right. \nl \left. - 1.81\cdot 10^{-3}\mathcal{C}^\text{con} + 10.0\mathcal{C}^F\right)
\end{align}
at linear order in the $\mathcal{C}^\circ_\circ$ terms.
We conclude that the solution is not under control for $\mathcal{C}^\circ_\circ=\mathcal{O}(1)$.

\subsection{A Bound on Perturbative Control}

We now analyze whether controlled dS vacua exist in other regions of the parameter space of the model. We will keep the same CY orientifold and brane data and assume $A_s=1$ as before. It is furthermore convenient to trade $W_0$ for the uplift parameter $\alpha$ using \eqref{alpha}. The adjustable parameters of the model are then $\alpha$, $g_s$, $K$ and $M$, and the subspace of the parameter space yielding dS vacua is simply its restriction to $\alpha\in]1,\frac{9}{4}[$.

As explained in Section \ref{pert}, we can estimate the degree of control at a candidate dS vacuum with the $\lambda_i$ parameters defined in \eqref{lambda00}--\eqref{lambda0}.
To this end, we write them as functions of $\alpha$, $g_s$, $K$ and $M$ using \eqref{tb0}--\eqref{q0}
(with the replacements $\xi\to\xi-\Delta\xi$ and $A_s\to A_s \e^{a_s\chi_s/24g_s}$). We furthermore substitute \eqref{parameters} and $A_s=1$.
This yields
\begin{align}
\lambda_1(\alpha,g_s) &= \text{max}\left(\frac{10+\alpha+4\alpha^2}{30|\alpha-1|}, \frac{135-16\alpha+16\alpha^2}{120|\frac{9}{4}-\alpha|} \right) \frac{2^{1/3}\pi g_s}{3^{1/3}(7\zeta(3))^{2/3}}, \label{lambda10} \\
\lambda_2(\alpha,g_s,K,M) &= \text{max}\left[\frac{1}{|\alpha-1|} \left|7-\frac{15}{\sqrt{9-\frac{64\pi c^\prime c^{\prime\prime}}{g_sM^2}}}\right|,\frac{1}{|\frac{9}{4}-\alpha|} \left|10-\frac{15}{\sqrt{9-\frac{64\pi c^\prime c^{\prime\prime}}{g_sM^2}}}\right| \right] \nl \times
\frac{2^{1/3} 3^{2/3} 256 \pi^6 c^{\prime 2} g_s^4 M^2 \alpha}{45(7\zeta(3))^{2/3} \left(3-\sqrt{9-\frac{64\pi c^\prime c^{\prime\prime}}{g_sM^2}}\right)} \nl \times \e^{-\frac{(126\zeta(3))^{2/3}}{3\pi g_s}+\frac{\pi}{12g_s} - \frac{1}{6} - \frac{8\alpha}{15}
+\frac{4\pi K}{3g_sM}-\frac{1}{6}\sqrt{9-\frac{64\pi c^{\prime}c^{\prime\prime}}{g_sM^2}} }, \label{lambda20} \\
\lambda_3(\alpha,g_s,K,M) &= \text{max}\left(\frac{20}{|\alpha-1|},\frac{55}{|\frac{9}{4}-\alpha|} \right) \frac{ 4096 \pi^{10} c^{\prime 2} g_s^2 KM \alpha^2}{525 \zeta(3) \left(3-\sqrt{9-\frac{64\pi c^\prime c^{\prime\prime}}{g_sM^2}}\right)^2}
\nl \times \e^{-\frac{2(126\zeta(3))^{2/3}}{3\pi g_s} +\frac{\pi}{6g_s} + \frac{2}{3} - \frac{16\alpha}{15}
+\frac{16\pi K}{3g_sM}-\frac{2}{3}\sqrt{9-\frac{64\pi c^{\prime}c^{\prime\prime}}{g_sM^2}}}, \label{lambda30} \\
\lambda_4(\alpha,g_s,M) &= \text{max}\left(\frac{1}{|\alpha-1|},\frac{1}{|\frac{9}{4}-\alpha|} \right) \frac{\alpha}{2(g_sM)^2} \left(1+ \frac{3}{\sqrt{9-\frac{64\pi c^\prime c^{\prime\prime}}{g_sM^2}}}\right), \label{lambda40} \\
\lambda_5(\alpha,g_s,K,M) &= \text{max}\left(\frac{32}{|\alpha-1|},\frac{88}{|\frac{9}{4}-\alpha|} \right) \frac{2^{1/3}3^{2/3} 625 (7\zeta(3))^{1/3} \left(3-\sqrt{9-\frac{64\pi c^\prime c^{\prime\prime}}{g_sM^2}}\right)^4 }{24461180928 \pi^{10} c^{\prime 4} g_s^{5} \alpha^4 } \nl \times \e^{ \frac{(126\zeta(3))^{2/3}}{\pi g_s}-\frac{\pi}{4g_s} - 2 + \frac{8\alpha}{5}
-\frac{32\pi K}{3g_sM}+\frac{4}{3}\sqrt{9-\frac{64\pi c^{\prime}c^{\prime\prime}}{g_sM^2}}}, \label{lambda50}
\end{align}
where $c^\prime =1.18$, $c^{\prime\prime}=1.75$ and we ignored subleading terms in $g_s$.

Our goal is to determine how much perturbative control is possible for dS vacua in this model, i.e., how small the $\lambda_i$ parameters can be made.
To simplify the analysis, we first bound each $\lambda_i$ from below by $\lambda_i\ge \hat \lambda_i \equiv \lambda_i(\hat\alpha_i)$, where we denote by $\hat\alpha_i$ the value of $\alpha\in\, ]1,\frac{9}{4}[$ for which $\lambda_i$ is minimized.
We find
\begin{align}
\hat \alpha_1 &= \frac{(3311+495\sqrt{82})^{1/3}}{12}-\frac{209}{12(3311+495\sqrt{82})^{1/3}}+\frac{2}{3} \approx 1.44, \notag \\
\hat \alpha_2 &= \frac{9}{4} - \frac{5}{4} \frac{\left|10\sqrt{9-\frac{64\pi c^\prime c^{\prime\prime}}{g_sM^2}}-15\right|}{\left|7\sqrt{9-\frac{64\pi c^\prime c^{\prime\prime}}{g_sM^2}}-15\right|+\left|10\sqrt{9-\frac{64\pi c^\prime c^{\prime\prime}}{g_sM^2}}-15\right|}, \notag \\
\hat \alpha_3 &= \frac{4}{3}, \qquad \hat \alpha_4 = \frac{13}{8}, \qquad \hat \alpha_5 = \frac{43}{16}-\frac{\sqrt{409}}{16}\approx 1.42. \label{hatalpha}
\end{align}

It is furthermore useful to define $\hat\lambda$ as the largest of the $\hat\lambda_i$ parameters at a given point in the parameter space:
\begin{align}
\hat\lambda(g_s,K,M)\equiv \sup&\left\{\hat\lambda_1\left(g_s\right),\hat\lambda_2\left(g_s,K,M\right),\hat\lambda_3\left(g_s,K,M\right),\hat\lambda_4\left(g_s,M\right),\hat\lambda_5\left(g_s,K,M\right)\right\}.
\end{align}
If dS vacua exist with $\lambda_i\ll 1$, all unknown corrections can self-consistently be neglected (aside from the $\mathcal{C}_s^\text{log}$, $\mathcal{C}^\xi_i$ corrections, which would only be negligible if in addition \eqref{c-cond} holds) and the problems described in Section \ref{problems} are avoided. According to our above definitions, $\lambda_i\ll 1$ implies $\hat\lambda_i\ll 1$ and thus $\hat\lambda\ll 1$. However, we will see that dS vacua with this property do not exist. Instead, $\hat\lambda \gtrsim 1$ in \emph{every} dS vacuum that exists in the parameter space.
To see this, we also define
\begin{align}
\lambda^{[K,M]}_\text{min} &\equiv \inf \left\{\hat\lambda(g_s,K,M) \left| g_s>0 \right.\right\}, \\
\lambda_\text{min} &\equiv  \inf \left\{\lambda^{[K,M]}_\text{min}\left| K,M\in\mathbb{N}, KM\le 149 \right.\right\}.
\end{align}
Hence, $\lambda^{[K,M]}_\text{min}$ denotes the smallest possible $\hat\lambda$ for a fixed flux choice $K$, $M$. Furthermore, $\lambda_\text{min}$ is the smallest possible $\hat\lambda$ in the full parameter space, i.e., scanning over all possible values of $g_s$, $K$ and $M$. By definition of $\lambda_\text{min}$, every dS vacuum in the model of \cite{Crino:2020qwk} has at least one $\lambda_i$ parameter satisfying $\lambda_i \ge \lambda_\text{min}$. We will therefore take $\lambda_\text{min}$ as an indicator of how well-controlled dS vacua can in principle be in the LVS.

\begin{figure}[t]
\centering
\includegraphics[width=0.4\linewidth]{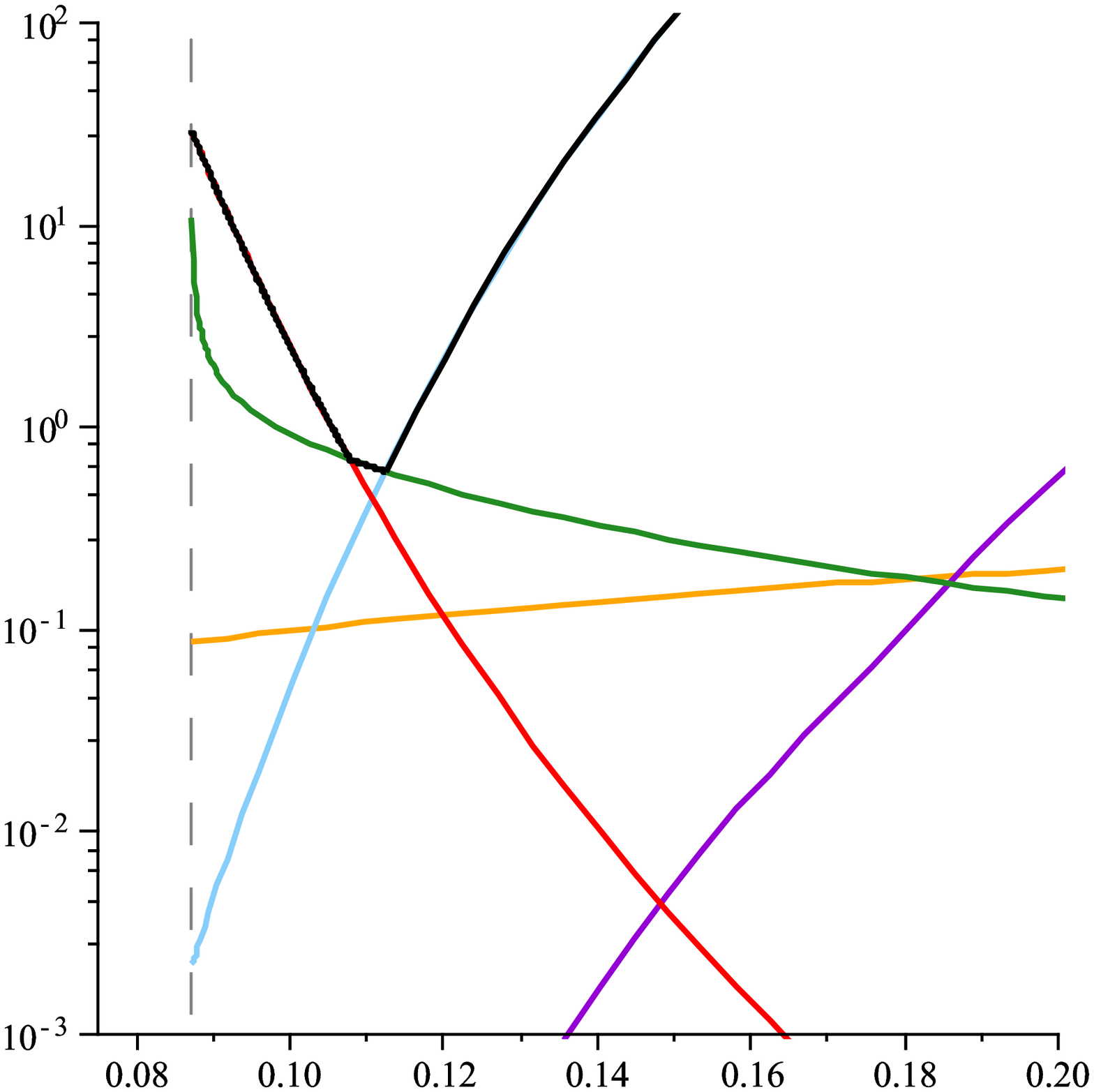} \qquad\qquad \includegraphics[width=0.4\linewidth]{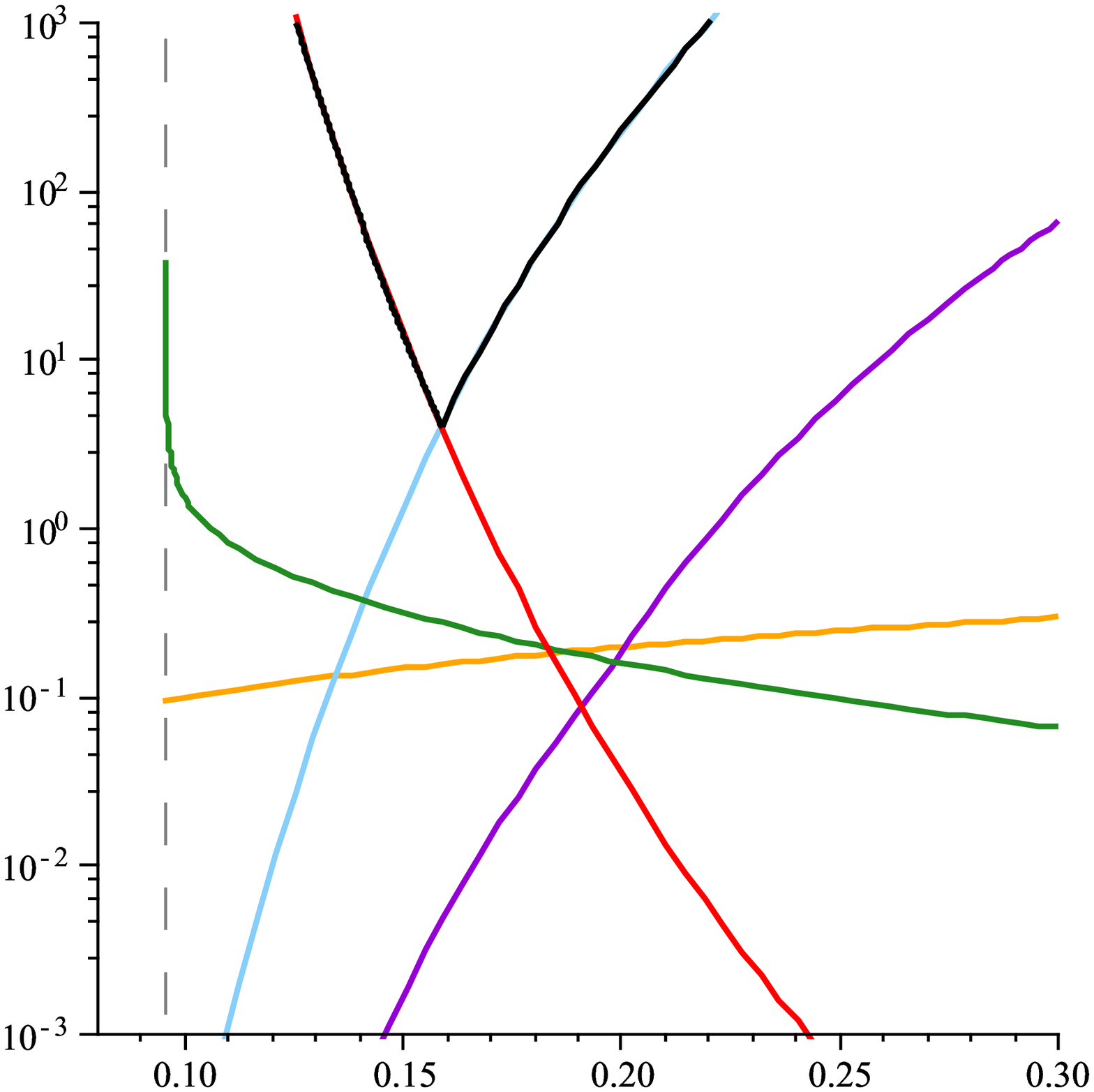}
\put(-413,90){$\hat\lambda_i$}
\put(-318,-8){$g_s$}
\put(-190,90){$\hat\lambda_i$}
\put(-95,-8){$g_s$}

\vspace{1em}

\includegraphics[width=0.4\linewidth]{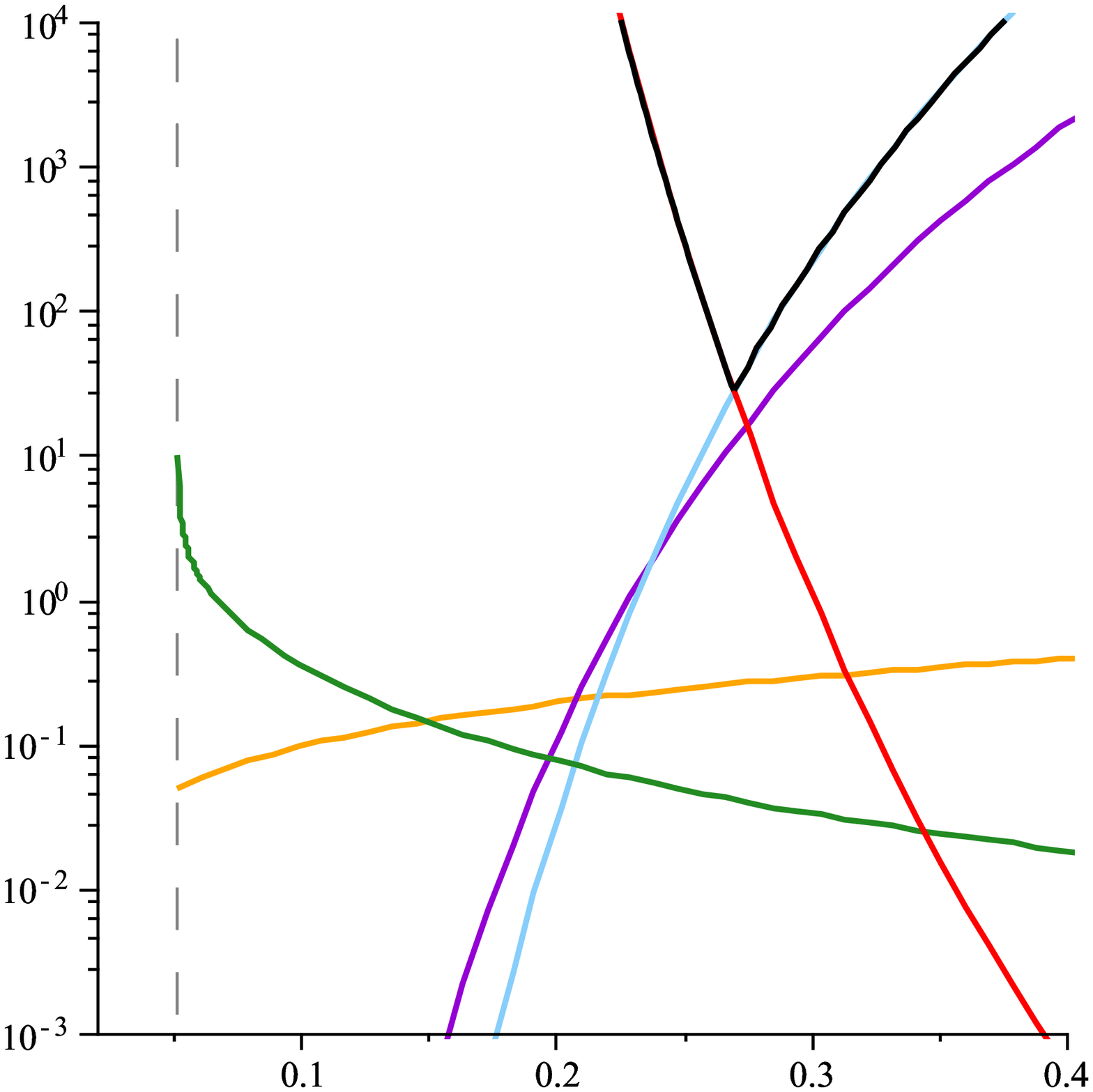} \qquad\qquad \includegraphics[width=0.4\linewidth]{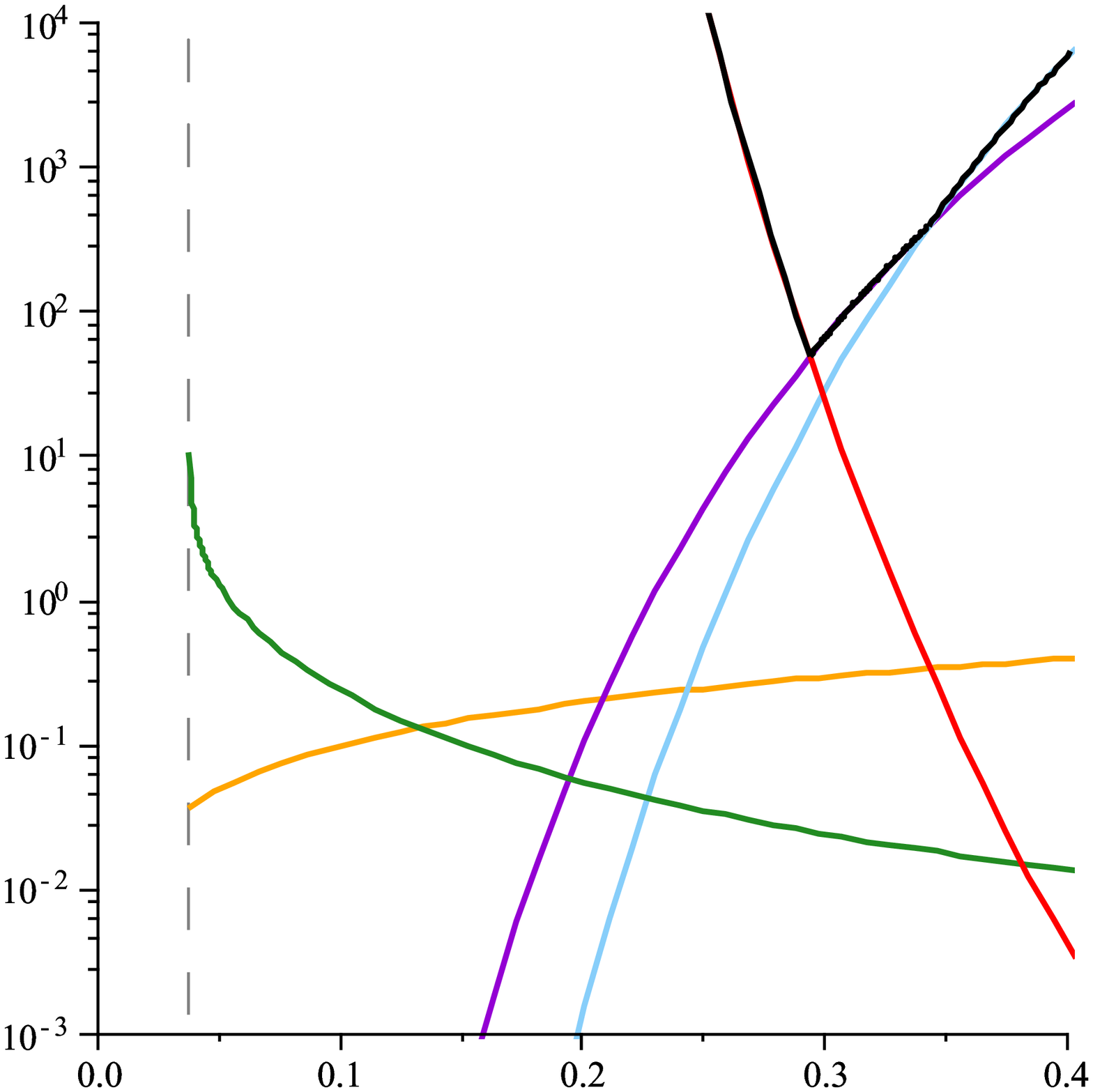}
\put(-413,90){$\hat\lambda_i$}
\put(-318,-8){$g_s$}
\put(-190,90){$\hat\lambda_i$}
\put(-95,-8){$g_s$}

\caption{Parameters $\hat\lambda_1$ (orange), $\hat\lambda_2$ (purple), $\hat\lambda_3$ (blue), $\hat\lambda_4$ (green) and $\hat\lambda_5$ (red) for the flux choices $K=5$, $M=23$ (upper left), $K=4$, $M=22$ (upper right), $K=2$, $M=30$ (lower left) and $K=1$, $M=35$ (lower right). The black curves denote $\hat\lambda$ and their minima yield $\lambda^{[K,M]}_\text{min}$. The grey dashed lines denote the conifold-instability bound $g_s=46.1/M^2$.\\
}
\label{lambda_bounds-1}
\end{figure}

To obtain $\lambda_\text{min}$, we first compute $\lambda^{[K,M]}_\text{min}$ for every allowed flux choice $K$, $M$.
This is straightforward using the expressions \eqref{lambda10}--\eqref{hatalpha}.
For example, consider $K=5$, $M=23$. As shown in Fig.~\ref{lambda_bounds-1}, $\hat\lambda$ is then given by $\hat\lambda_3$, $\hat\lambda_4$ or $\hat\lambda_5$, depending on our choice for $g_s$. As is evident from the figure, the smallest $\hat\lambda$ is obtained for the value $g_s=0.112$ where $\hat\lambda_3$ and $\hat\lambda_4$ intersect.
We thus find $\lambda^{[5,23]}_\text{min} = \hat\lambda_3(0.112,5,23)= 0.605$.

\begin{figure}[t]
\centering
\includegraphics[width=0.4\linewidth]{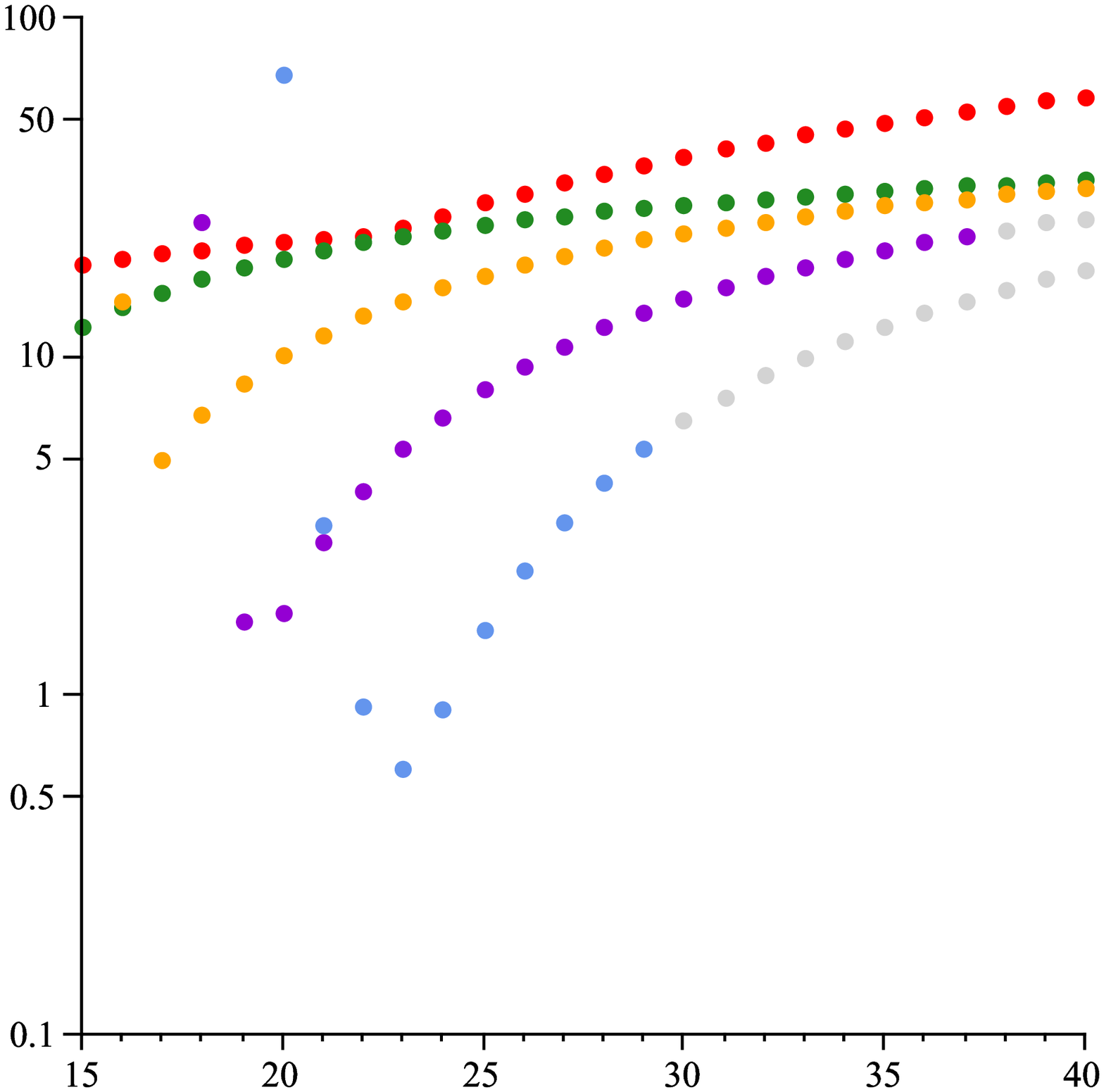}
\put(-208,90){$\lambda^{[K,M]}_\text{min}$}
\put(-95,-10){$M$}
\caption{$\lambda^{[K,M]}_\text{min}$ for $K=1$ (red), $K=2$ (green), $K=3$ (orange), $K=4$ (purple) and $K=5$ (blue). Grey points exceed the tadpole bound $KM\le 149$.
The lowest colored point corresponds to $\lambda_\text{min}$.\\}
\label{lambda_bounds-2}
\end{figure}

Analogously, we can compute $\lambda^{[K,M]}_\text{min}$ for all other $K$, $M$ compatible with the tadpole condition $KM\le 149$ and thus determine $\lambda_\text{min}$ (cf.~Fig.~\ref{lambda_bounds-2}). The parameter choice with the smallest $\lambda^{[K,M]}_\text{min}$ turns out to be the example we just gave, i.e., $K=5$, $M=23$.
We thus find
\begin{equation}
\lambda_\text{min} =\lambda^{[5,23]}_\text{min} = 0.605. \label{bl}
\end{equation}
We have thus shown that every dS vacuum in this model has at least one $\lambda_i$ parameter $\ge 0.605$. Note that the actual $\lambda_i$ parameters are even larger than this. Indeed, the solutions do not saturate our lower bound since we derived it by substituting a different $\alpha=\hat\alpha_i$ in each $\lambda_i$.
Also note that the above flux choice uses up most of the available tadpole. Restricting to smaller $K$, $M$ in order to leave more room for bulk fluxes that stabilize the dilaton and the complex-structure moduli would further increase $\lambda_\text{min}$.

An example for a dS vacuum where the largest $\lambda_i$ is close to the bound is
\begin{equation}
g_s=0.112,\qquad W_0=29,\qquad K=5,\qquad M=23,
\end{equation}
corresponding to
\begin{equation}
\mathcal{V} = 5.24\cdot 10^6, \qquad \tau_s = 13.6, \qquad \zeta= 3.40\cdot 10^{-6}, \qquad \alpha = 1.47.
\end{equation}
This yields
\begin{equation}
\lambda_1 = 0.115, \qquad \lambda_2 = 6.53 \cdot 10^{-5}, \qquad \lambda_3=0.738, \qquad \lambda_4 =0.739, \qquad \lambda_5=0.323.
\end{equation}
The result for $\tau_s$, $V_0$ and $m_3^2$ including the linear $\mathcal{C}^\circ_\circ$ terms is
\begin{align}
\tau_s &= 13.6-0.238\mathcal{C}^\text{KK}_s+6.56\cdot 10^{-5} \mathcal{C}^\text{KK}_b+13.5 \mathcal{C}^\text{log}_s +23.8 \mathcal{C}^\xi_1 + 1.77\mathcal{C}^\xi_2+ 0.368\mathcal{C}^\text{flux}\nl +8.80\cdot 10^{-2}\mathcal{C}^\text{con} + 0.161\mathcal{C}^F, \\
V_0 &= 6.84\cdot 10^{-20}\left( 1-2.62\cdot 10^{-2}\mathcal{C}^\text{KK}_s+5.90\cdot 10^{-5} \mathcal{C}^\text{KK}_b+3.51 \mathcal{C}^\text{log}_s +4.42 \mathcal{C}^\xi_1 \right. \nl \left. + 6.22\cdot 10^{-2} \mathcal{C}^\xi_2 -0.406\mathcal{C}^\text{flux}+0.668\mathcal{C}^\text{con} - 0.178\mathcal{C}^F\right), \\
m_3^2 &= 5.49\cdot 10^{-19}\left( 1+ 1.96\cdot 10^{-2}\mathcal{C}^\text{KK}_s-7.62\cdot 10^{-6} \mathcal{C}^\text{KK}_b-4.51 \mathcal{C}^\text{log}_s -4.97 \mathcal{C}^\xi_1 \right. \nl \left. +7.78\cdot 10^{-2} \mathcal{C}^\xi_2  +0.835 \mathcal{C}^\text{flux}-0.499\mathcal{C}^\text{con} + 0.365\mathcal{C}^F\right).
\end{align}
For $\mathcal{C}^\circ_\circ=\mathcal{O}(1)$, this is clearly problematic. We stress again that it is not possible to get significantly better controlled dS solutions anywhere in the parameter space because of \eqref{bl}.
We have thus shown in general that the model only admits dS vacua in an uncontrolled regime with large corrections.

We close this section with a few remarks on possible ways to avoid the above conclusions. As stated before, we cannot exclude that the numerical coefficients $\mathcal{C}^\circ_\circ$ are smaller than our expectation $\mathcal{C}^\circ_\circ=\mathcal{O}(1)$ in particular models, e.g., due to conspiracies involving powers of $\pi$ or specific cancellations. However, there does not seem to be a reason to expect such a property for all coefficients.
For example, we already computed one contribution to the $\mathcal{C}^\text{con}$ coefficient in \eqref{itau} and found it to be $-1.97$. In order for $|\mathcal{C}^\text{con}|$ to be small or zero, this number would have to cancel very precisely with further $\mathcal{O}(1)$ contributions to $\mathcal{C}^\text{con}$ from other $\alpha^\prime$ corrections. Such a cancellation would be quite miraculous.

One might hope that another way out is to consider different values for the constants $\gamma_0$, $\Lambda_0$ and $A_s$, which we discussed in Section \ref{setup} and later set to 1 following the literature. Note that these constants cannot be chosen freely but are fixed by the geometry and the stabilization of the complex-structure moduli in a given model. One can furthermore check that all three constants (if not set to 1 as above) show up with positive powers in some of the $\lambda_i$ and with negative powers in others. We therefore expect that, say, a very small $\gamma_0\ll 1$ would not significantly change the bound we derived above.

Finally, our result should not be qualitatively affected by the caveat discussed in Section \ref{setup} related to the off-shell $Z$ dependence,
as most of the corrections we considered are not sensitive to the latter.

\section{Conclusions}
\label{concl}

In this paper,
we systematically studied various types of corrections to the LVS potential in a setup with two K\"ahler moduli, a conifold modulus and a nilpotent superfield describing the anti-brane uplift. Some of these corrections were already derived prior to this work, but only in the case without the conifold modulus and the uplift \cite{Cicoli:2007xp, Cicoli:2008va, Berg:2007wt, Conlon:2010ji}. Moreover, we worked out several corrections to the potential that have not been considered
in the context of the LVS before.

We then derived analytic expressions for the leading corrections to the moduli vevs, the vacuum energy and the moduli masses.
These results
are fully general and may be useful for a variety of future studies of the LVS.
A key issue we identified is that corrections can appear less suppressed in these expressions than in the off-shell potential. This is due to the fact that the $\tau_s$ vev scales like $1/g_s$ in the LVS and a related effect we called the non-perturbative no-scale structure (NPNS).

A common lore is
that LVS vacua are extremely well-controlled because of their exponentially large volumes. However, our results show that this is not the case.
In particular, the $\mathcal{C}_s^\text{log}$ and $\mathcal{C}^\xi_i$ corrections are not suppressed by any small parameters in the relevant expressions. Even worse, the $\mathcal{C}_s^\text{log}$ and $\mathcal{C}^\xi_1$ terms in the $\tau_s$ vev scale like $1/g_s$ and thus blow up at small coupling.
We argued that this implies an exponential uncertainty in the volume and the uplift parameter $\alpha$ and potentially affects the signs of the vacuum energy and the moduli masses.
In addition, we showed in an explicit CY model previously studied in \cite{Crino:2020qwk}
that further types of corrections are unsuppressed at every point in the parameter space admitting dS vacua (assuming that the unknown coefficients $\mathcal{C}^\circ_\circ$ are $\mathcal{O}(1)$ numbers).
These results suggest
that it is impossible to construct controlled LVS dS vacua, i.e., dS vacua where all unknown corrections can be self-consistently neglected.

While the part of our analysis in Section \ref{dsmin} focussed on the explicit model of \cite{Crino:2020qwk},
it is straightforward to derive analogous bounds on the $\lambda_i$ parameters in any other model by substituting the corresponding CY and brane data into our general equations.
It would be very interesting to check whether there are geometries for which the bound on the $\lambda_i$ is significantly weaker or
whether this can be ruled out in general.
In principle, it would also be important
to make progress on explicit computations of the various coefficients $\mathcal{C}^\circ_\circ$ in smooth CYs. However, this is a formidable task that may not be feasible
in the near future.

In any case, our work reinforces the point that it is extremely hard to construct explicit dS vacua in string theory and that a loss of control
seems inevitable in any attempt to do so. In view of the intense recent activity in the field, we will hopefully be able to tell soon
whether this reflects a fundamental inconsistency of dS space or just our inability to find it within the vast string landscape.

Finally, our results are also relevant for the stability of non-supersymmetric AdS solutions in the LVS. While most of the corrections we studied are less dangerous or absent in the AdS case, this is not true for all of them. In particular, the expressions in Section \ref{problems} are general and remain valid in the AdS case. In the simplest setup without anti-branes or a conifold region, we can ignore the $\mathcal{C}_b^\text{KK}$, $\mathcal{C}^\text{flux}$ and $\mathcal{C}^\text{con}$ terms in these equations and set $\alpha=0$ in the remaining terms. As is evident from the expressions for the vevs and masses of the K\"ahler moduli, the $\mathcal{C}_s^\text{log}$ and $\mathcal{C}_i^\xi$ corrections are then still unsuppressed and could thus, together with the $\mathcal{C}^F$ terms, potentially destabilize the AdS vacua.

\section*{Acknowledgments}

I would like to thank Severin L\"ust, Christoph Mayrhofer, Jakob Moritz, Andreas Schachner, Gary Shiu and Timo Weigand for useful discussions/correspondence.

\bibliographystyle{utphys}
\bibliography{groups}

\end{document}